\documentclass[iop, revtex4]{emulateapj}
\usepackage{times}
\usepackage{lscape}

\usepackage{graphicx,bm,amssymb}
\usepackage{ulem}
\usepackage{epsfig}
\usepackage{amssymb}
\usepackage{natbib}
\usepackage{pstricks}
\usepackage{psfrag}
\usepackage[colorlinks=true,linkcolor=blue,citecolor=blue]{hyperref}
\def\araa{ARA\&A}
\def\apj{ApJ}
\def\apjl{ApJ}
\def\apjs{ApJS}
\def\aap{A\&A}
\def\mnras{MNRAS}
\def\nat{Nature}
\def\procspie{Proc.~SPIE}

\newcommand{\be}{\begin{equation}}
\newcommand{\ee}{\end{equation}}
\newcommand{\bary}{\begin{eqnarray}}
\newcommand{\eary}{\end{eqnarray}}

\shorttitle{ Optical polarimetric and multiwavelength flaring activity..}
\shortauthors{Fraija, N. et al.}
\begin{document}
\title{Optical polarimetric and multiwavelength flaring activity of blazar 3C\,279}
\author{N. Fraija$^{1\dagger}$, E. Ben{\'i}tez$^1$,  D. Hiriart$^2$, M. Sorcia$^1$, J. M. L\'opez$^3$,  R. M\'ujica$^4$,   J. I. Cabrera$^5$, and  A. Galv\'an-G\'amez$^1$}%
\affil{$^1$ Instituto de Astronom\' ia, Universidad Nacional Aut\'onoma de M\'exico, Circuito Exterior, C.U., A. Postal 70-264, 04510 Cd. de M\'exico,  M\'exico\\
$^2$ Instituto de Astronom\'ia, Universidad Nacional Aut\'onoma de M\'exico, Ensenada, Baja California, Mexico\\
$^3$Facultad de Ciencias, Universidad Aut\'onoma de Baja California, Campus El Sauzal, Ensenada B.C., Mexico\\
$^4$ Instituto Nacional de Astrof\'isica, \'Optica y Electr\'onica, Apdo. Postal 51 y 216, 72000 Tonantzintla, Puebla, Mexico\\
$^5$ Facultad de Ciencias, Universidad Nacional Aut\'onoma de M\'exico, Apdo. Postal 70-264, 04510 Cd. de M\'exico, Mexico\\
}
\email{$\dagger$nifraija@astro.unam.mx}
 \date{\today} 
\begin{abstract}
An exhaustive analysis of 9-year optical R-band photopolarimetric data of the flat-spectrum radio quasar 3C\,279 from 2008 February 27 to 2017 May 25  is presented, alongside with multiwavelength observing campaigns performed during the flaring activity exhibited in 2009 February/March, 2011 June, 2014 March/April, 2015 June and 2017 February.   In the R-band, this source showed the maximum brightness state of $13.68\pm 0.11$ mag ($1.36\pm0.20$ mJy) on 2017 March 02, and  the lowest brightness state ever recorded of  $18.20\pm 0.87$ mag ($0.16\pm0.03$ mJy) on 2010 June 17.    During the entire period of observations, the polarization degree varied between $0.48\pm0.17$\% and $31.65\pm0.77$\% and the electric vector position angle exhibited large rotations between $82.98^\circ \pm0.92$ and $446.32^\circ \pm1.95$.  Optical polarization data show that this source has a stable polarized component that varied from $\sim$6\% (before the 2009 flare) to $\sim$13\% after the flare.  The overall behavior of our polarized variability data supports the scenario of jet precessions as responsible of  the observed large rotations of the electric vector position angle.     Discrete correlation function analysis show that the lags between gamma-rays and X-rays compared to the optical R-band fluxes are $\Delta t \sim$ 31\,d and $1$\,d in 2009. Lags were also found among gamma-rays compared with X-rays and radio of  $\Delta t \sim$ 30\,d and $43$\,d in 2011, and among radio and optical-R band of $\Delta t \sim$ 10\,d in 2014.  A very intense flare in 2017 was observed in optical bands with a dramatic variation in the polarization degree (from $\sim$ 6\% to 20 \%) in 90 days without exhibiting flaring activity in other wavelengths. 
\end{abstract}

\section{Introduction}
%
%
Blazars, a subclass of radio loud active galactic nuclei (AGN) that launches ultra-relativistic jets that points near the observer's line of sight \citep{1979ApJ...232...34B}, are known to display  large variability in all spectral bands. 
Blazars are commonly divided in BL Lac objects and Flat Spectrum Radio Quasars (FSRQ) based on the equivalent width (EW) of the optical emission lines. \citep[see][]{1996MNRAS.281..425M,2012agn..book.....B}. One of the main characteristics of blazars is its high and variable polarization emission, that have been observed from the radio wavebands up to the optical bands.

Multiwavelength flux variations  have been observed in timescales from minutes \citep{1995ARA&A..33..163W} to years \citep[see, e.g.,][]{1988ApJ...325..628S}.  In the optical bands, polarization variations have been observed also in timescales ranging from some minutes \citep[see, e.g.][]{2015ApJ...809L..27B, 2017ApJS..232....7F} and also,  large rotations of the electric vector position angle (EVPA) have been reported during flaring activities \citep[see, e.g.][]{2014ApJ...794...54S,2016A&A...590A..10K}. Optical polarization variability studies play an important role in the establishment of the intensity and direction of the magnetic field within the emitting region \citep[see, e.g.][]{2013ApJS..206...11S,2014ApJ...794...54S,2017ApJS..232....7F}. New facilities like the Imaging X-ray Polarimetry Explorer \citep[IXPE;][]{2016SPIE.9905E..17W} and with the All-sky Medium Energy Gamma-ray Observatory \citep[AMEGO;][]{2019arXiv190304607R} will allow  to perform multiwavelength polarization variability studies in blazars.

Located at a redshift of z=0.536 \citep{1965ApJ...142.1667L},  the FSRQ 3C\,279 is one of the brightest and more variable extragalactic sources in the $\gamma$-ray sky. This object was the first FSRQ in being detected by EGRET \citep[an instrument onboard  satellite Compton gamma-ray observatory;][]{1992ApJ...385L...1H, 1999ApJS..123...79H, 2001ApJ...553..683H} and  a ground-based atmospheric Cherenkov telescope in very-high energies \citep[VHE; E$>$100 GeV;][]{2018ATel11239....1N}.  Due to the large optical EVPA rotations in coincidence with flaring activity in $\gamma$-rays, the object 3C\,279 has been subject to long-term monitoring programs in order to understand the mechanism responsible of the VHE emission.   Due to different temporal and spectral behavior in high- and low- activity states, the broadband spectral energy distribution (SED) of 3C\,279  is usually modeled with leptonic and hadronic models.  In the leptonic scenario, the external Compton (EC) and one-zone synchrotron self-Compton (SSC) models have been invoked \citep{2012ApJ...754..114H, 2009ApJ...703.1168B}. In the hadronic scenario the proton-synchrotron radiation and  secondaries $e^{\pm}$-synchrotron  emission producing in the pion decay products have been considered \citep{2013ApJ...768...54B, 2017MNRAS.467L..16P}.\\ 

Studies of polarization degree and the EVPA in 3C\,279 have been done in both flaring and quiescent states. During the flaring activity from 2008 November to 2009 March, multiwavelength observations showing EVPA variability coincident with gamma-ray variations were reported by \cite{2010Natur.463..919A}. This result shows a   clear optical/$\gamma$-ray correlation. A shock compression of an ordered helical magnetic field in an axisymmetric jet model  \citep{2015ApJ...804...58Z}  and the {\rm bent} jet model \citep{2010IJMPD..19..701N}  were proposed to interpret this atypical correlations.  Similarly, the {\rm bent} jet model was required to explain the EVPA variations  presented in the multi-wavelength campaign in 2011 \citep{2014A&A...567A..41A}.   In addition,  \cite{2008A&A...492..389L} presented a multi-wavelength study based on the X-ray, optical and radio bands and polarimetric observations collected during 2006-2007. They reported large EVPA variations which were explained by a large-scale helical magnetic field moving through the jet \citep{2008Natur.452..966M}. \cite{1999ApJ...512..157K} presented the first far-infrared (FIR) polarization results obtained with \textit{ISOPHOT} between 1996 and 1997. They found variability of the FIR polarization without any variation of the observed flux. This behavior was interpreted by a disturbance of the magnetic field around two emitting regions.  In radio wavebands,  \cite{2018arXiv180504588R} used high-frequency radio interferometry (VLBI) polarization imaging to investigate the magnetic field's topology in the emitting region.

Recent simulations have been done in order to explain the physics in the jet of blazars \citep[e.g.][]{2019arXiv190312381B}. Specifically, for blazar 3C\,279 they found correlations among the variability observed in the gamma-rays (GeV) with optical bands, and among radio and X-rays.  \cite{2018ApJ...856...99P} and  \cite{2018arXiv180704046B}  studied the SED and the variability behavior around the Flare observed in June 2015.  On one hand, \cite{2018ApJ...856...99P} found that the multiwavelenth behavior challenges the one-zone leptonic models and on the other hand,  \cite{2018arXiv180704046B} found that the lepto-hadronic models are favored over the one-zone SSC models. Finally, in order to find evidence for a precession jet and a double-jet structure in blazar 3C\,279, \cite{2019A&A...621A..11Q} investigated the parsec-scale kinematics of the superluminal components observed in this blazar. These authors propose that a double-jet structure scenario in 3C\,279 properly describe the observations.
  
In this work,  we analyze the long-term optical R-band photopolarimetric data  of FSRQ 3C\,279 from 2008 February 27 to 2017 May 25, along with multiwavelength observing campaigns performed during the flaring activity exhibited in 2009 February/March, 2011 June, 2014 March/April, 2015 June and 2017 February.   The paper is arranged as follows: in Section 2, the optical R-band photopolarimetric data and statistical analysis per cycles are presented. In Section 3  a description with the analysis done of the multi-wavelength data are shown. Section 4 shows the modeling of the variable multiwavelength spectral energy distribution and finally, conclusions in Section 5 are presented.

\section{Optical R-band photopolarimetric data and Statistical Analysis}

The optical R-band photopolarimetric data reported in this paper were carried out with the 0.84 m f/15 Ritchey-Chr\'etien telescope at the Observatorio Astron\'omico Nacional of San Pedro M\'artir (OAN-SPM) in Baja California, Mexico\footnote{http://www.astrossp.unam.mx/blazars.}.  A single beam polarimeter so-called POLIMA\footnote{http://haro.astrossp.unam.mx/blazars/instrument/instrument.html} was used.  Details on calibration,  data reduction, the correction due to  host galaxy contribution and the conversion between R-band magnitudes and fluxes are shown in \cite{2013ApJS..206...11S,2014ApJ...794...54S,2017ApJS..232....7F}.

\begin{figure*}
\centering
\includegraphics[width=0.95\textwidth]{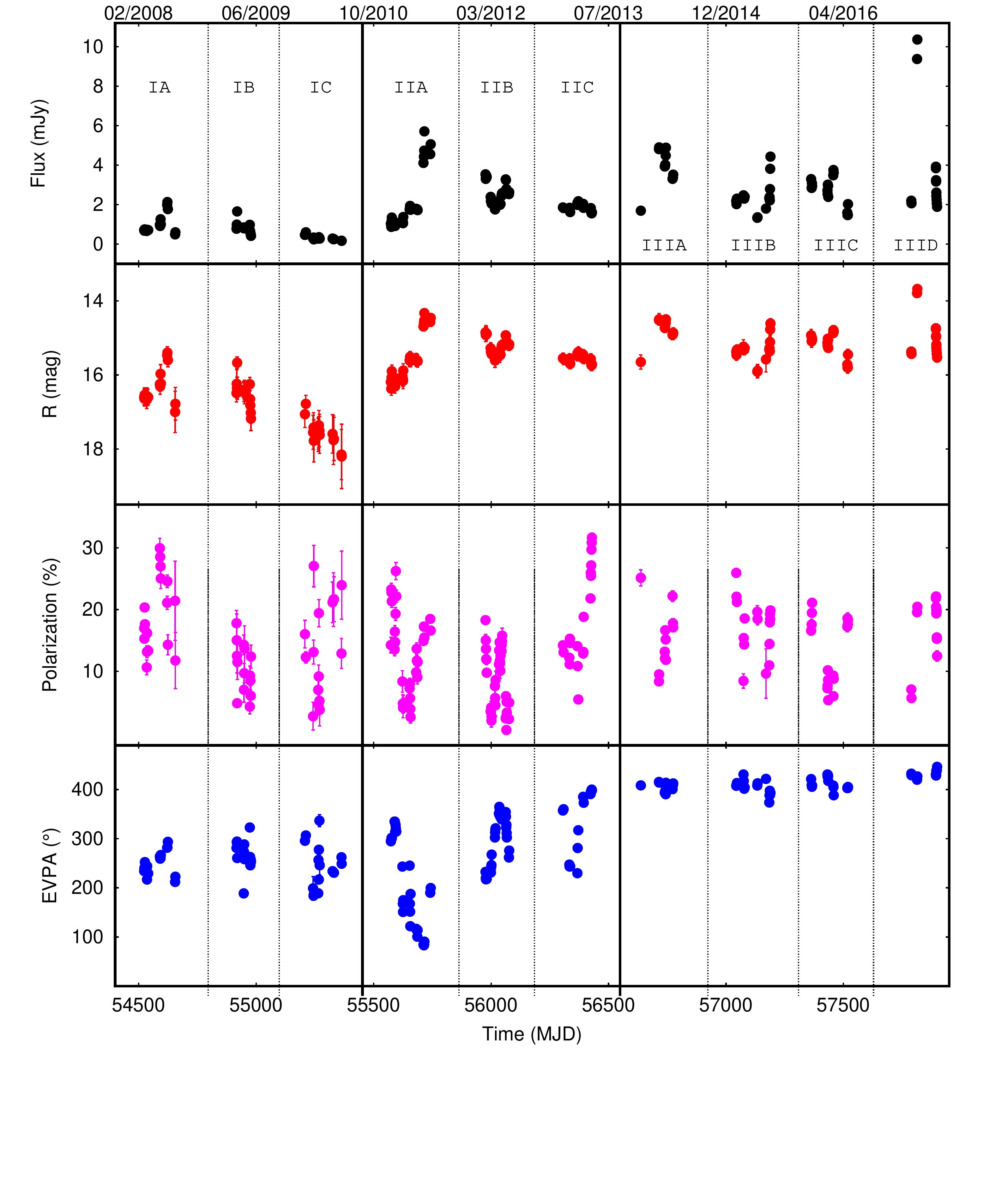}
\caption{OAN-SPM photometric R-band light curve of 3C\,279 obtained from 2008 February 27 (MJD 54523) to 2017 May 25 (MJD 57898) including the polarimetric variability of P(\%) and EVPA. From top to bottom: R-band mag, P(\%) and the EVPA variations are shown.  Vertical solid and dashed lines separate the monitoring period into cycles I, II and III and their corresponding sub-cycles A, B, C and D, respectively.}
\label{optical_all}
\end{figure*}

Figure \ref{optical_all} shows the long-term R-band photopolarimetric light curves of the FSRQ 3C\,279, for nine years of observations (data are available online, see Table \ref{all_data}). These light curves  were built with 188 data points collected during the period  2008 February 27 (MJD 54523) to 2017 May 25 (MJD 57898).  From the top, panels show the optical R-band flux, magnitude, polarization degree and EVPA.   The maximum and minimum  brightness states are: R$=13.68\pm 0.11$ mag ($10.36\pm  0.20$ mJy) and  R$=18.20\pm 0.87$ mag ($0.16\pm  0.03$ mJy), and  were detected in 2017 March 02 and 2010 June 17,  respectively.  During this long-term period, 3C\,279 exhibited an average polarization degree of $\sim$ 14\% with a preferred EVPA direction of $\sim311^\circ$.  The low variations  of  the polarization degree during the large rotations have been  explained in the literature as due to symmetry of the toroidal components of the helical magnetic field \citep{2008Natur.452..966M, 2016MNRAS.462.4267J, 2010ApJ...710L.126M}, and also due to turbulent magnetic field resulting in a random direction of the polarization vector \citep{1985ApJ...290..627J, 2014ApJ...780...87M}.\\
 Considering the entire data set, no significant correlations were found among photopolarimetric data and in the normalized Stokes parameters $q=Q/F_R=P\cos2\theta$ and $u=U/F_R=P\sin2\theta$.  During the long-term monitoring period, 3C\,279 showed in some periods or cycles several flares.   
 \subsection{Statistical Analysis per cycle}
%
The optical polarization angle exhibits large variable rotations during the nine years of observation.  In order to analyze the long-term R-band photopolarimetric data, the entire observations are divided in three cycles (I, II and III). The optical EVPA in cycle I presents random variations,  in cycle II it exhibits dramatic swings with remarkable maximums and minimums, and finally, in cycle III the EVPA shows a constant tendency.  These results agree with the results recently reported  in \citep{2019A&A...626A..78B}.

In order to search correlations per cycle among the optical flux, polarization degree, EVPA and normalized Stokes parameters, a statistical analysis was done (see Table \ref{stat_cicle}). In addition,
the Pearson's correlation coefficients along with the $p$ values were estimated and reported in Table \ref{pearson_year}.   The amplitude of the variations $Y$(\%), the fluctuation index $\mu$ and  the fractional variability index $\cal F$ reported in Table \ref{stat_cicle} are calculated using the equations \citep{1996A&A...305...42H}.    
 \begin{equation}
Y(\%) = \frac{100}{\cal h S i}\sqrt{(S_{\rm max}-S_{\rm min})^2-2\sigma^2_c} \;\; ,
\end{equation}
\begin{equation}
\mu = 100\frac{\sigma_S}{\cal hSi}\% \; ,
\end{equation}
and
\begin{equation}
{\cal F} = \frac{S_{\rm max}-S_{\rm min}}{S_{\rm max}+S_{\rm min}} \;,
\end{equation}
respectively, with $S_{\rm max/min}$ and ${\cal hSi}$ the maximum/minimum and average values of the optical flux, the polarization degree and the EVPA and $\sigma_c =\sqrt{\sigma^2_{\rm max}+\sigma^2_{\rm min}}$.  In addition, the timescale of flux variations are calculated by 

\begin{equation}\label{var}
\tau_{\nu}= \mid \frac{\Delta t }{\Delta \ln F}  \mid\,,
\end{equation}

where $\Delta t$ is the time interval between two adjacent optical fluxes $F_i$ and $F_j$, with $i, j=1,...,M-1$ and $M$ the number of data points.

\subsubsection{\sf Cycle I}
\begin{figure*}
\centering
\includegraphics[width=0.95\textwidth]{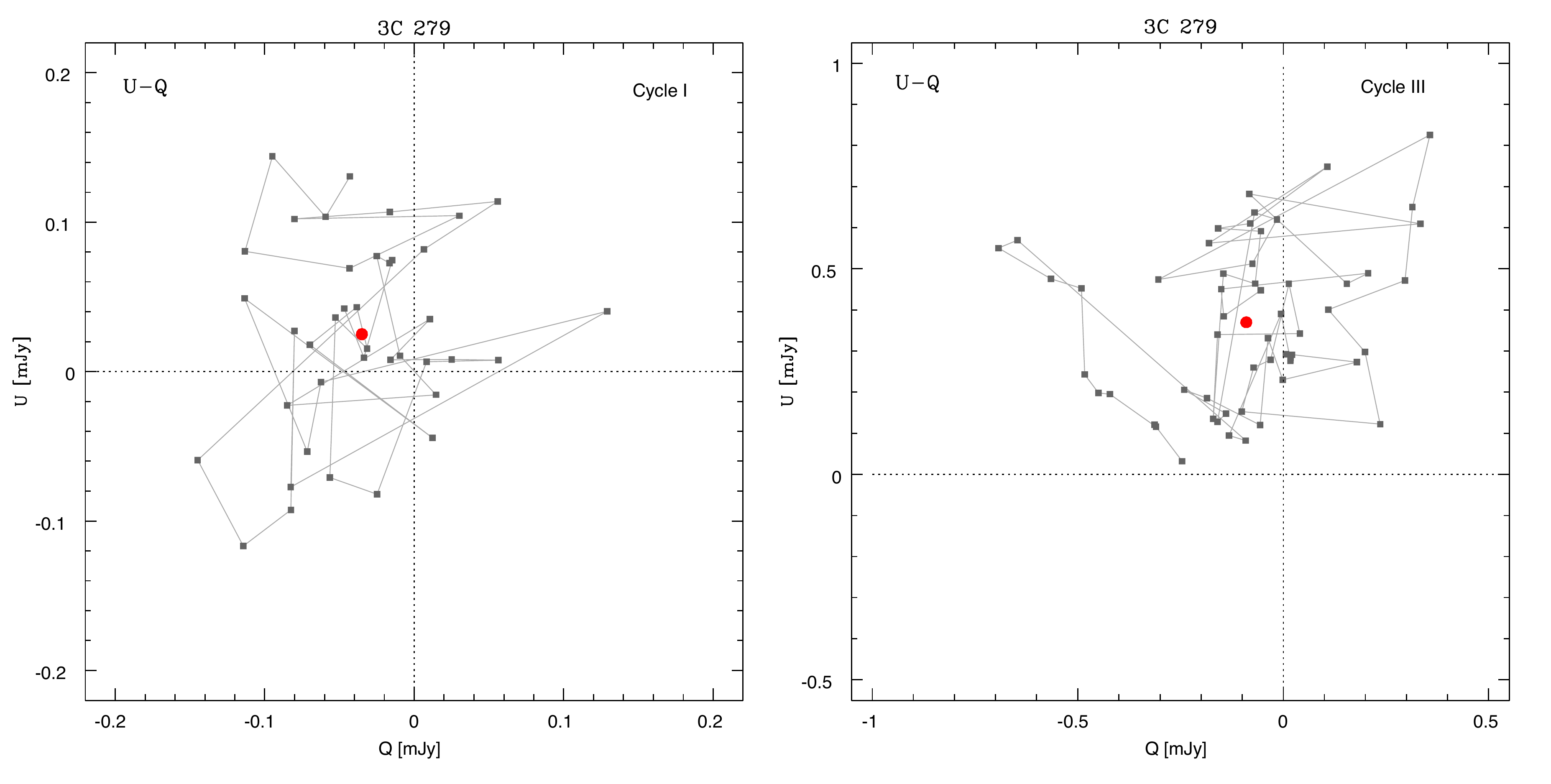}
\caption{Q - U absolute Stokes parameter planes for cycles I  and III. The red points correspond to the obtained mean constant values and show that a stable polarization component exist in both cycles, see text.}
\label{polarization_cycles}
\end{figure*}

\begin{figure*}
\centering
\includegraphics[width=1.\textwidth]{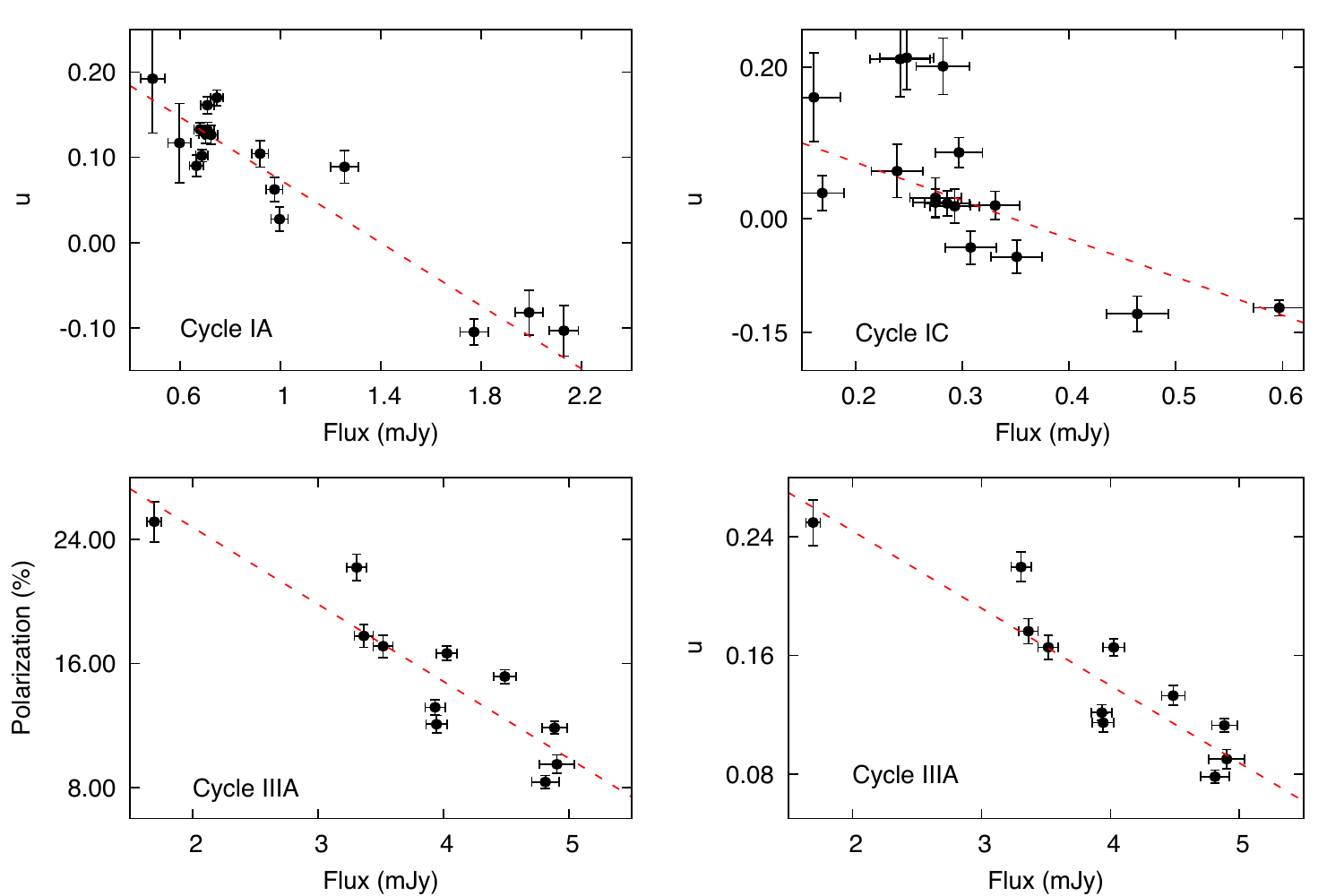}
\caption{Optical R-band photopolarimetric correlations found in different cycles.  Left: optical flux vs the normalized Stokes parameter $u$ found in cycle IA (top panel)  and optical flux vs P(\%) found in cycle IIIA (bottom panel).   Right:  optical flux vs the normalized Stokes parameter $u$ found in cycle IC (top panel) and IIIA  (bottom panel).}
\label{correlations}
\end{figure*}

 Cycle I covers the period 2008 February 27 (MJD 54523) to 2010 June 17 (MJD 55364).   During this cycle, the EVPA had a variation of  $\sim156^\circ$ with the lowest average flux of $0.71\pm0.01$ mJy.  The optical flux, the polarization degree and  the EVPA reached their maximum values of 2.13 mJy, 29.94\% and 336.76$^\circ$, respectively in timescales of $\Delta t\sim$ 2 years.
Using the method presented in \cite{2013ApJS..206...11S},  the absolute Stokes parameters Q and U are plotted and shown in the left panel of Figure \ref{polarization_cycles}. The red point shows the obtained mean constant value, and its position indicates that a stable polarization component is present. The mean absolute Stokes parameters estimated for this component are:  $\langle Q\rangle=-0.05\pm0.01$ and  $\langle U\rangle=0.03\pm 0.01$.  The polarization degree and its dispersion found for the stable component are  $P_c=6.83\pm1.43$ and $\sigma_p=8.7\%$, respectively.  The value of  the EVPA for the stable component is $\theta_c=67.7^\circ\pm 2.5$. For clarity, cycle I is further divided in three sub-cycles:

\paragraph{\sf Sub-cycle IA: From 2008 February 27 (MJD 54523) to 2008 July 09 (MJD 54656).}
During this sub-cycle, the optical R-band magnitudes, the polarization degree  and  the EVPA values exhibited variations of up to  0.26 mag, 10\%  and 28$^\circ$ in a single day.  The minimum variability timescale found is $4.98\pm0.62$ days.   The  analysis showed that the optical flux and the Stokes parameter ${\rm u}$ are strongly correlated.   This correlation is shown in Figure \ref{correlations} (upper left-hand panel).   The best-fit coefficients  of the linear function are  ${\rm m_{\rm Fu}}=-0.18\pm 0.01$ and ${\rm b_{\rm Fu}}=0.26\pm 0.01$ with a chi-square $\chi^2=6.92$.\\

\paragraph{\sf Sub-cycle IB: From 2009 March 26 (MJD 54916) to 2009 May 27 (MJD 54978).}
During this sub-cycle, the optical R-band magnitude, the polarization degree  and  the EVPA exhibited variations of up to  0.78 mag, 8\%  and 60$^\circ$ in a single day.  The minimum variability timescale is $1.49\pm0.09$ days. No significant correlation was found between the  photopolarimetric data and the Stokes parameters.
 
\paragraph{\sf Sub-cycle IC: From 2010 January 11 (MJD 55207) to 2010 June 17 (MJD 55364).}
 During this sub-cycle, the optical R-band magnitude, the polarization degree  and  the EVPA displayed variations of up to 0.20 mag, 14\%  and 40$^\circ$ in a single day.  The minimum variability timescale is $5.90\pm0.59$ days.  The  analysis showed that the optical flux and Stokes parameter ${\rm u}$ is strongly correlated.   This correlation is shown in Figure \ref{correlations} (upper right-hand panel). The best-fit coefficients  of the lineal function are ${\rm m_{Fq}}=-0.50\pm 0.04$ and ${\rm b_{Fq}}=0.18\pm 0.02$ with a chi-square $\chi^2=6.04$.

\subsubsection{\sf Cycle II}
Cycle II the period 2012 March 12 (MJD 55998) to 2013 May 17 (MJD 56429).  During this cycle, the EVPA oscillate between 83$^\circ$  and 400$^\circ$, exhibiting a very large variation of  $\sim317^\circ$ and the lowest average polarization degree of $12.51\pm0.10$ \%.   The optical flux increased up to 5.71 mJy in a timescale of $\Delta t\sim$ 0.4 years, the polarization degree went up to  31.65\% in a timescale of $\Delta t=$ 1 year and the EVPA reached the value of 318$^\circ$ in a timescale of $\Delta t\sim$ 2 years. Cycle II is further divided in three sub-cycles:

\paragraph{\sf Sub-cycle IIA: From 2011 January 12 (MJD 55573) to 2011 July 01 (MJD 55743).}  During this sub-cycle, the optical R-band magnitude, the polarization degree and the EVPA exhibited variations of up to 0.30 mag, 7\%  and 78$^\circ$ in one day.   The minimum variability timescale is $3.49\pm0.17$ days.

\paragraph{\sf Sub-cycle IIB: From 2012 February 19 (MJD 55976) to 2013 May 17 (MJD 56077).} During this sub-cycle, the optical R-band magnitude, the polarization degree  and  the EVPA  exhibited variations of up to 0.16 mag, 4\%  and 21$^\circ$ in one day.  The minimum variability timescale is $12.94\pm0.55$ days. No significant correlation was found between the  photopolarimetric data and the Stokes parameters. 

\paragraph{\sf Sub-cycle IIC: From 2013 January 13 (MJD 56305) to 2013 May 17 (MJD 56429).}  The highest polarization value of $31.65\pm0.77$ \% was detected.   The optical R-band magnitude, the polarization degree  and  the EVPA  exhibited variations of up to 0.08 mag, 4\%  and 52$^\circ$ in one day.  The minimum variability timescale is $29.81\pm0.99$ days.   

\subsubsection{\sf Cycle III}
Cycle III the period 2014 April 27 (MJD 56774) and 2017 May 25 (MJD 57898).   During this cycle, the EVPA shows a tendency towards a constant value.  The optical flux reached the maximum value of 10.36 mJy in a timescale of $\Delta t\sim$ 2 years, the polarization degree of 25.93\% in a timescale of $\Delta t=$ 1 year and EVPA of 446.32$^\circ$ in a timescale of $\Delta t\sim$ 2 years.  As in cycle I, the absolute Stokes parameters Q and U are plotted and shown in the right panel of Figure \ref{polarization_cycles}. The red point shows the obtained mean constant value, and again its position indicates that a stable polarization component is also present in cycle III. The mean absolute Stokes parameters estimated for this component are:  $\langle Q\rangle=-0.09\pm0.01$ and  $\langle U\rangle=0.37\pm 0.02$.  The polarization degree and its dispersion found for the stable component are  $P_c=13.30\pm0.56$ and $\sigma_p=6.4\%$, respectively.  The value of  the EVPA for the stable component is $\theta_c=52.6^\circ\pm 2.1$. It is worth noting that the EVPA in the R-band  is consistent with the results reported by \cite{2018ApJ...858...80R}.  These authors measured an EVPA of $\sim 50^\circ$ and direction of the jet  projected in the sky of $-135^\circ$. Cycle III is further divided in four sub-cycles.

\paragraph{\sf Sub-cycle IIIA: From 2013 December 11 (MJD 56637) to 2014 April 28 (MJD 56775).}  During this sub-cycle, the optical R-band magnitude, the polarization degree  and  the EVPA  exhibited variations of up to 0.09 mag, 4\%  and 23$^\circ$ in one day.  The minimum variability timescale is $11.61\pm0.33$ days.   The  analysis showed that the optical flux vs Stokes parameter $u$,  and  the optical flux vs polarization degree are strongly anti-correlated.   These correlations are shown in Figure \ref{correlations} (lower panels). The best-fit coefficients of the lineal function are ${\rm m_{Fp}}=-4.97\pm 0.27$ and ${\rm b_{Fp}}=34.70\pm 1.15$ with a chi-square $\chi^2=19.22$ (left-hand panel) and  ${\rm m_{Fq}}=-0.05\pm 0.01$ and ${\rm b_{Fq}}=0.35\pm 0.01$ with a chi-square $\chi^2=13.55$ (right-hand panel).
\paragraph{\sf Sub-cycle IIIB: From 2015 January 21 (MJD 57043) to 2015 June 16 (MJD 57189).}   During this sub-cycle, the optical R-band magnitude, the polarization degree  and  the EVPA  exhibited variations of up to 0.34 mag, 7\%  and 15$^\circ$ in one day.  The minimum variability timescale is $3.17\pm0.12$ days.  No significant correlation was found between the  photopolarimetric data and the Stokes parameters. 
 \paragraph{\sf Sub-cycle IIIC: From 2015 December 06 (MJD 57362) to 2016 May 11 (MJD 57519).}   During this sub-cycle, the optical R-band magnitude, the polarization degree  and  the EVPA  exhibited variations of up to 0.35 mag, 4\%  and 17$^\circ$ in one day. The minimum variability timescale is $3.04\pm0.11$ days.   No significant correlation was found between the  photopolarimetric data and the Stokes parameters. 
 \paragraph{\sf Sub-cycle IIID: From 2017 February 04 (MJD 57788) to 2017 May 25 (MJD 57898).}  The highest brightness state of ($10.36\pm 0.20$) mJy was observed.     During this sub-cycle, the optical R-band magnitude, the polarization degree  and  the EVPA  exhibited variations of up to 0.21 mag, 4\%  and 12$^\circ$ in one day.  The minimum variability timescale is $2.90\pm0.08$ days.  No significant correlation was found between the  photopolarimetric data and the Stokes parameters. 
%
%
\section{Multi-wavelength Data Analysis}
\subsection{Multi-wavelength data}
In order to study our R-band data along with quasi-simultaneous multi-wavelength data around the flares observed in February/March 2009, June 2011,  March/April 2014,  June 2015 and February 2017,  the description of the data set will be given in this section. 
\paragraph{Gamma-ray observations.} Gamma-ray data collected by Fermi-LAT was collected between the energy range 0.1 - 300 GeV using the public database\footnote{http://fermi.gsfc.nasa.gov/ssc/data}.  
\paragraph{X-ray observations.}   The Swift-BAT/XRT  data used in this paper is publicly available and located in the HEASARC database\footnote{http://swift.gsfc.nasa.gov/cgi-bin/sdc/ql?}.   Data from the Proportional Counter Array  (XRTE - PCA) used was obtained following standard procedures through the text script in HEASOFT and the RXTE tools\footnote{http://heasarc.gsfc.nasa.gov/docs/xte/asm\_products.html}.  
\paragraph{Additional optical observations.}   Additional optical observations used in this paper were obtained by the Whole Earth Blazar Telescope which is part of the GLAST-AGILE Support Program (GASP-WEBT)\footnote{http://www.oato.inaf.it/blazars/webt/}, Small \& Moderate Aperture Research Telescope System \citep[SMART;][]{2012ApJ...756...13B},  University's Hard Labor Creek\footnote{http://www.astro.gsu.edu/HLCO/}, Khomas Highlands\footnote{https://www.lsw.uni-heidelberg.de/projects/hess/ATOM/} and  Hiroshima\footnote{http://hasc.hiroshima-u.ac.jp/telescope/kanatatel-e.html}.
\paragraph{Radio observations.} The radio data from the 40m Owens Valley Radio Observatory  \citep[OVRO;][]{2011ApJS..194...29R}  was collected as part of an ongoing long-term $\gamma$-ray blazar monitoring campaign. These data is publicly available\footnote{http://www.astro.caltech.edu/ovroblazars/}.  
\subsection{Flaring activity in 2009 February/March}
\begin{figure*}
\centering
\includegraphics[width=0.95\textwidth]{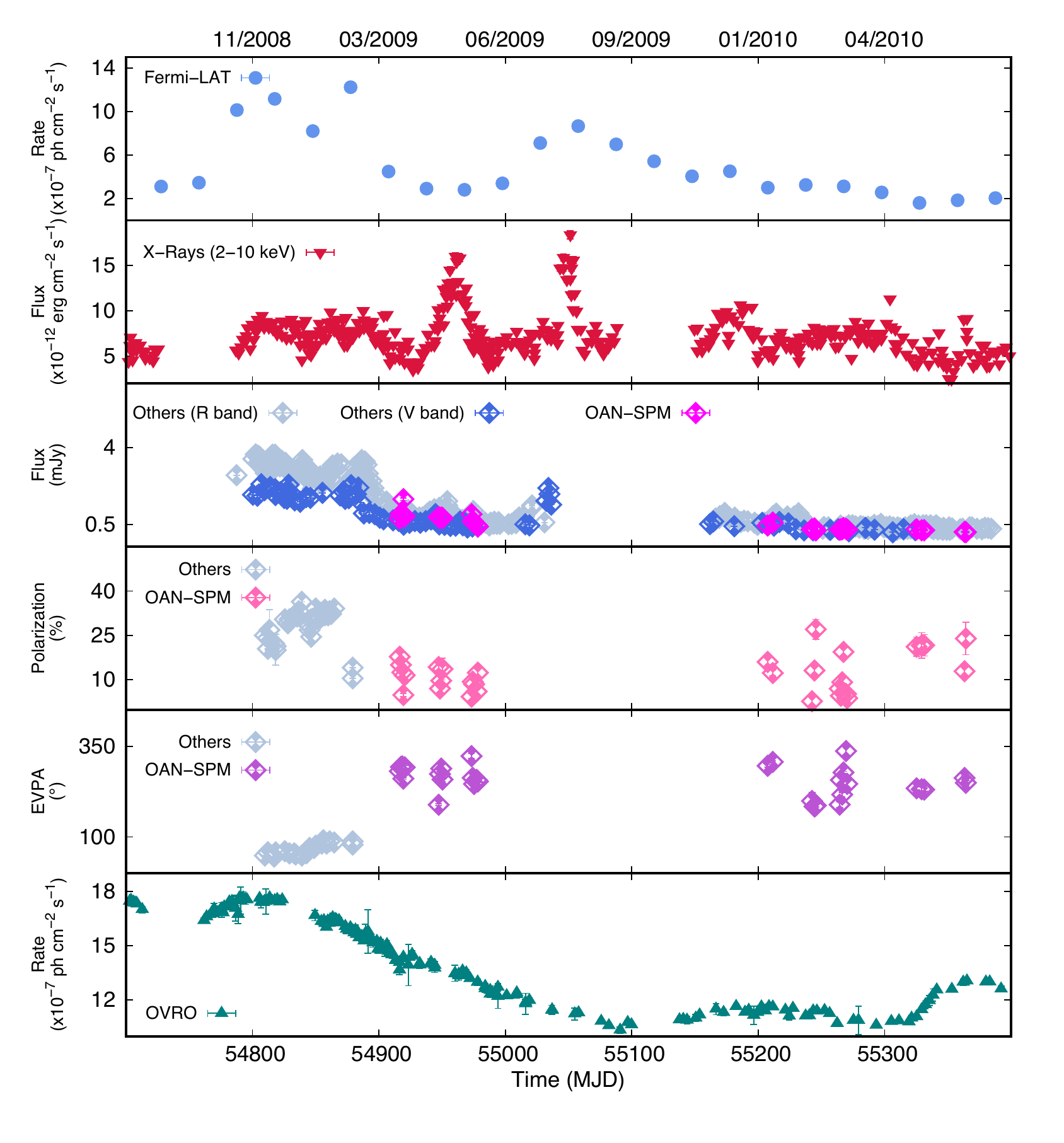}
\caption{Multiwavelength lightcurves of 3C\,279 between 2008 August  22 (MJD 54590) and 2010 July 23 (MJD 55400). They include (from top to bottom): $\gamma$-ray flux data above 100 MeV from Fermi-LAT, (2 - 10) keV X-ray flux data from Swift-XRT and XRTE-PCA, R-band fluxes obtained with POLIMA+84cm at the OAN-SPM and from the GASP-WEBT database:  P(\%) and EVPA from OAN-SPM and from Kanata and KVA databases, and radio flux OVRO databse.}
\label{flare:2009}
\end{figure*}

The multiwavelength lightcurves of 3C\,279 between 2008 August  22 (MJD 54590) and 2010 July 23 (MJD 55400) are presented in Figure  \ref{flare:2009}. They include (from top to bottom): $\gamma$-ray flux above 100 MeV collected by  Fermi-LAT, (2 - 10) keV X-ray flux (Swift-XRT and XRTE-PCA), optical fluxes in R band measured by OAN-SPM and GASP,  polarization degree and EVPA detected by OAN-SPM,  Kanata and KVA and radio flux (OVRO).\\
During the period between 2009 March 26 (MJD 54916) and 2010 June 17 (MJD 55364), the optical R-band magnitude, the polarization degree  and  the EVPA displayed variations of up 0.78 mag, 14\%  and 70$^\circ$ in one day.  The minimum variability timescale is $1.49\pm0.09$ days, which corresponds to a size of emitting radius $r_d=(5.03\pm0.03)\times 10^{16}\,{\rm cm}$. The  analysis of the optical flux and Stokes parameter ${\rm u}$ shows a strong correlation when  $\gamma$-ray, X-ray, optical and radio fluxes reached the quiescent level.\\ 

During the analyzed period,  3C\,279 showed diverse flaring events in $\gamma$-ray, X-ray, optical and radio bands accompanied by a large change of EVPA  and a random variation of the polarization degree.  A remarkable $\gamma$-ray/optical correlation together with a large variation of the EVPA (from 100$^\circ$ to 280$^\circ$) and change of the polarization degree during the high activity is observed. This correlation could be due to optical and $\gamma$-ray emissions that are spatially correlated, providing the presence of a highly ordered magnetic field. As suggested by  \cite{2015ApJ...804...58Z} and \cite{2010IJMPD..19..701N} a possible explanation is related with a shock compression of an ordered helical magnetic field in an axisymmetric and {\rm bent}  jet models, respectively.  Subsequently,  the X-ray light curve exhibited two flares: the first flare was detected on 2009 April 29 (MJD 54950) and $\sim$ 90 days later a second flare in 2009 July 28 was registered.  During the first X-ray flare, the source shows no variations in the $\gamma$-ray and optical R-band domain. During the second one, variations in the $\gamma$-rays and the R-band were observed. These behaviors suggest that the first X-ray flare could be generated by secondary $e^\pm$ pairs generated by the charged pion decay products, and the second one by a one-zone SSC model.\\

\begin{figure*}
\centering
\includegraphics[width=19 cm , height=21 cm]{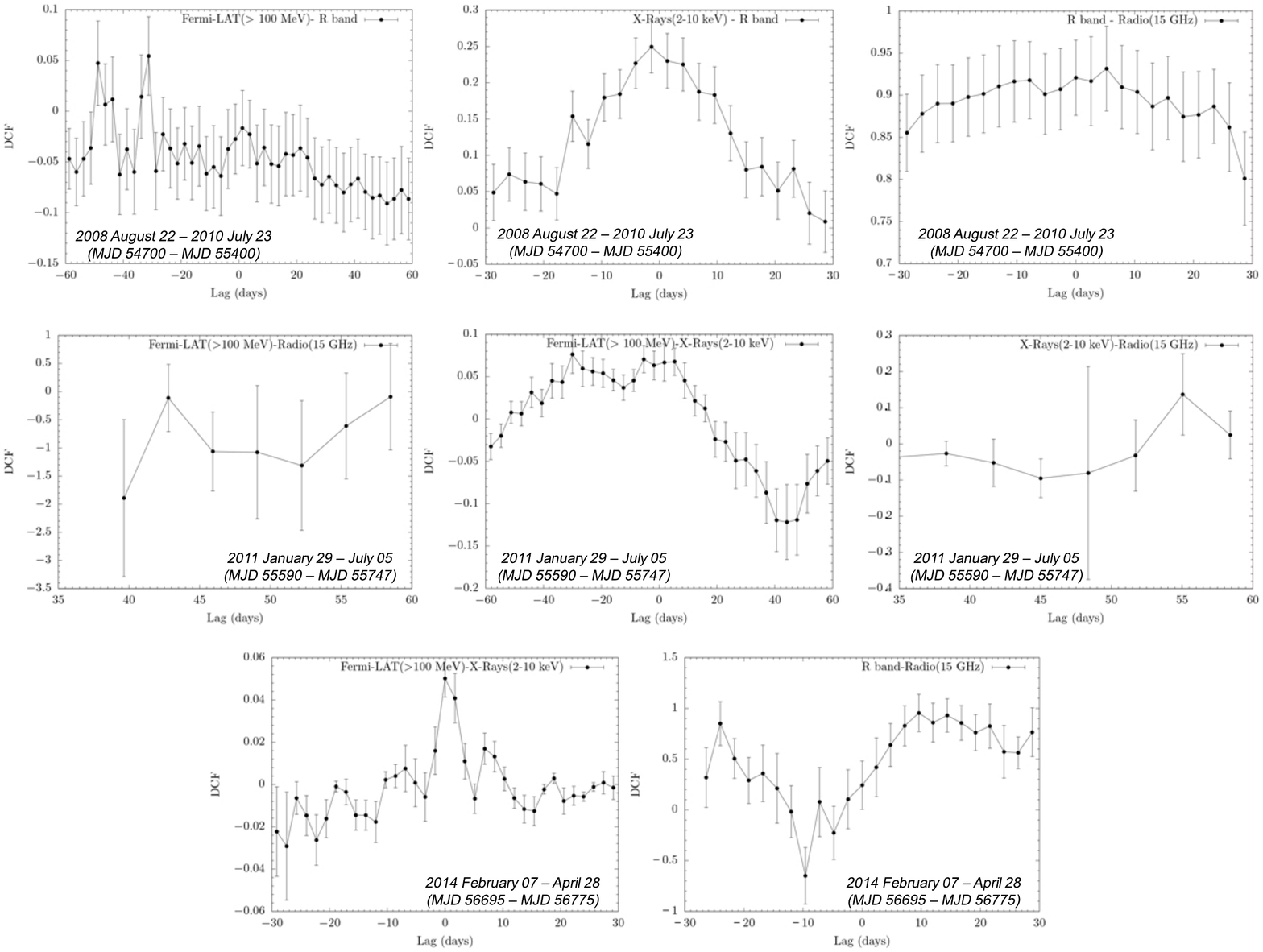}
\caption{Discrete correlation function (DCF) among the gamma-ray, X-ray, optical R-band and radio bands during the periods  2008 August 22 (MJD 54700) - 2010 July 23 (MJD 55400) (upper panels),  2011 January 29 (MJD 55590) -  July 05 (MJD 55747) (middle panels) and 2014 February 07 (MJD 56695) - April 28 (MJD 56775)  (lower panels).}
\label{dcf}
\end{figure*}

The discrete correlation function \citep[DCF; ][]{1988ApJ...333..646E} was calculated to quantify the correlation of the flux variations and to measure any possible lags among optical R-band, gamma-ray, X-ray and radio bands. Figure \ref{dcf} (upper panels) shows the DCF for almost two years between 2008 August 22 (MJD 54700) and 2010 July 23 (MJD 55400). 
The left-hand panel shows that the DCF of the gamma-ray and optical R-band fluxes have two peaks  at $\sim$ - 31 and $\sim$ - 49 days.  This shows that gamma-ray and optical bands are not correlated.   The middle- and right-hand panels show a lag among  X-rays and radio bands with respect to optical R-band of $\Delta t\sim$ 1.36 and 5.2 days, respectively, thus indicating that the X-ray, optical and radio bands are not correlated.   This analysis suggests that:  i) a contribution of secondary $e^\pm$ pairs radiating synchrotron photons from radio wavelengths to X-rays could be present.  These pairs are usually generated in the interactions between Fermi-accelerated protons with photons from synchrotron radiation, broad emission lines or infrared dust emission. In order to have these kind of process, extreme conditions such as a large magnetic field and a very compact emitting region are required,  ii) contributions from two electron populations, with one population  accelerated more efficiently than the other are required.  On one hand, the radio to X-ray emission needs lower energy electrons. On the other hand, the observed gamma-ray emission requires higher energy relativistic electrons .  

\subsection{Flaring activity in 2011 June}

\begin{figure*}
\centering
\includegraphics[width=0.95\textwidth]{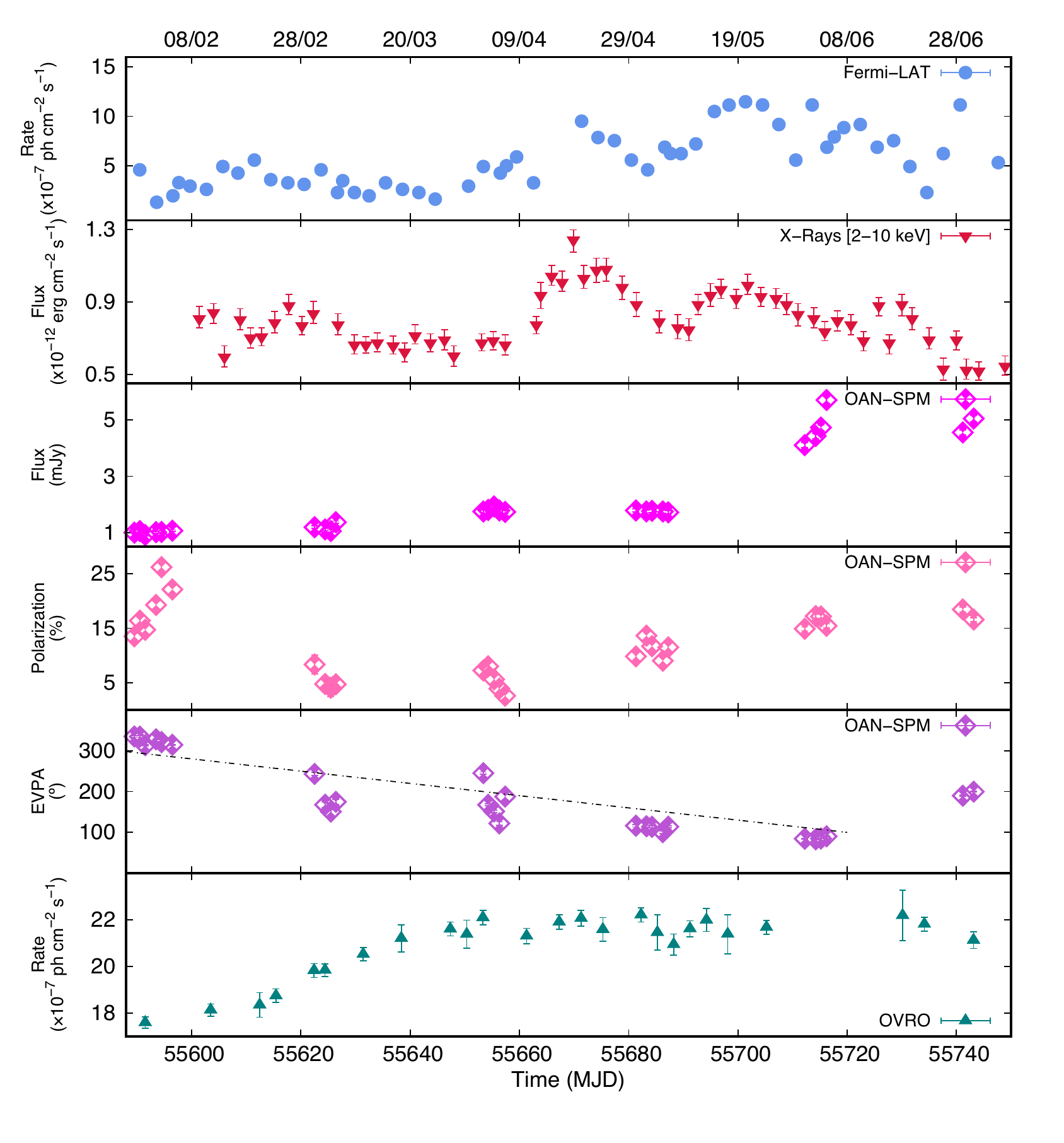}
\caption{Multiwavelength lightcurves of 3C\,279 obtained from 2011 January  29 (MJD 55590) and July 05 (MJD 55747) are shown. They include (from top to bottom): $\gamma$-ray flux data above 100 MeV from Fermi-LAT, (2 - 10) keV X-ray flux data from XRTE-PCA, optical flux, polarization degree and EVPA from OAN-SPM  and radio flux data from OVRO database.}
\label{flare:2011}
\end{figure*}

The multiwavelength lightcurves of 3C\,279 between  2011 January  29 (MJD 55590) and July 05 (MJD 55747)  are presented in Figure  \ref{flare:2011}. They include (from top to bottom): $\gamma$-ray flux above 100 MeV (Fermi-LAT), (2 - 10) keV X-ray flux collected by XRTE-PCA, optical flux, polarization degree and EVPA measured by OAN-SPM  and radio flux detected by OVRO.\\
%
During this period the optical flux,  the polarization degree and the EVPA displayed large variations in a timescale of $\Delta=$ 4 months.   The optical flux increases from 0.9 to 5.6 mJy and the polarization degree increases from 5\% to 20\%.  The EVPA swings, going from  294$^\circ$ to 90$^\circ$ and finally, goes back to 200$^\circ$.  The EVPA decreases gradually without showing significant variations around $\gamma$- and X-ray high activities presented at  April 19 (MJD 55670) and  May 19 (55700).  The black dotted-dashed line  shows the tendency of the EVPA before the highest $\gamma$-ray and optical flare at June 28 (55740). It indicates that during these activities 3C\,279 presents two emitting regions: one where $\gamma$-rays and  X-rays are spatially correlated and another where $\gamma$ and optical bands are spatially correlated.  It suggest that a node propagating along the helical magnetic-field lines up to the last flare is presented.\\

The DCF was calculated again to estimate possible lags among gamma-ray, X-ray and radio bands.   During the period between  2011 January  29 (MJD 55590) and July 05 (MJD 55747),  the resulting DCF between  gamma - radio  (left), gamma - X (middle) and X - radio (right) bands are shown in Figure \ref{dcf} (middle panels).    The DCF between gamma-ray - radio (left) and X-rays - radio (right) show similar peaks at 42.79 and 55.06 days.  It indicates  that radio wavelengths are not correlated with gamma-ray and X-ray bands.  The DCF between gamma-ray and X-ray bands shows a peak at 30.01 days,  which points out that both bands are also not correlated.   The previous analysis suggests possible contributions from two emitting regions, each one with different electron populations. One low electron population producing  radio wavelengths, and other higher radiating from X-ray to gamma-ray emission.
\subsection{Flaring activity in 2014 March/April}
\begin{figure*}
\centering
\includegraphics[width=0.95\textwidth]{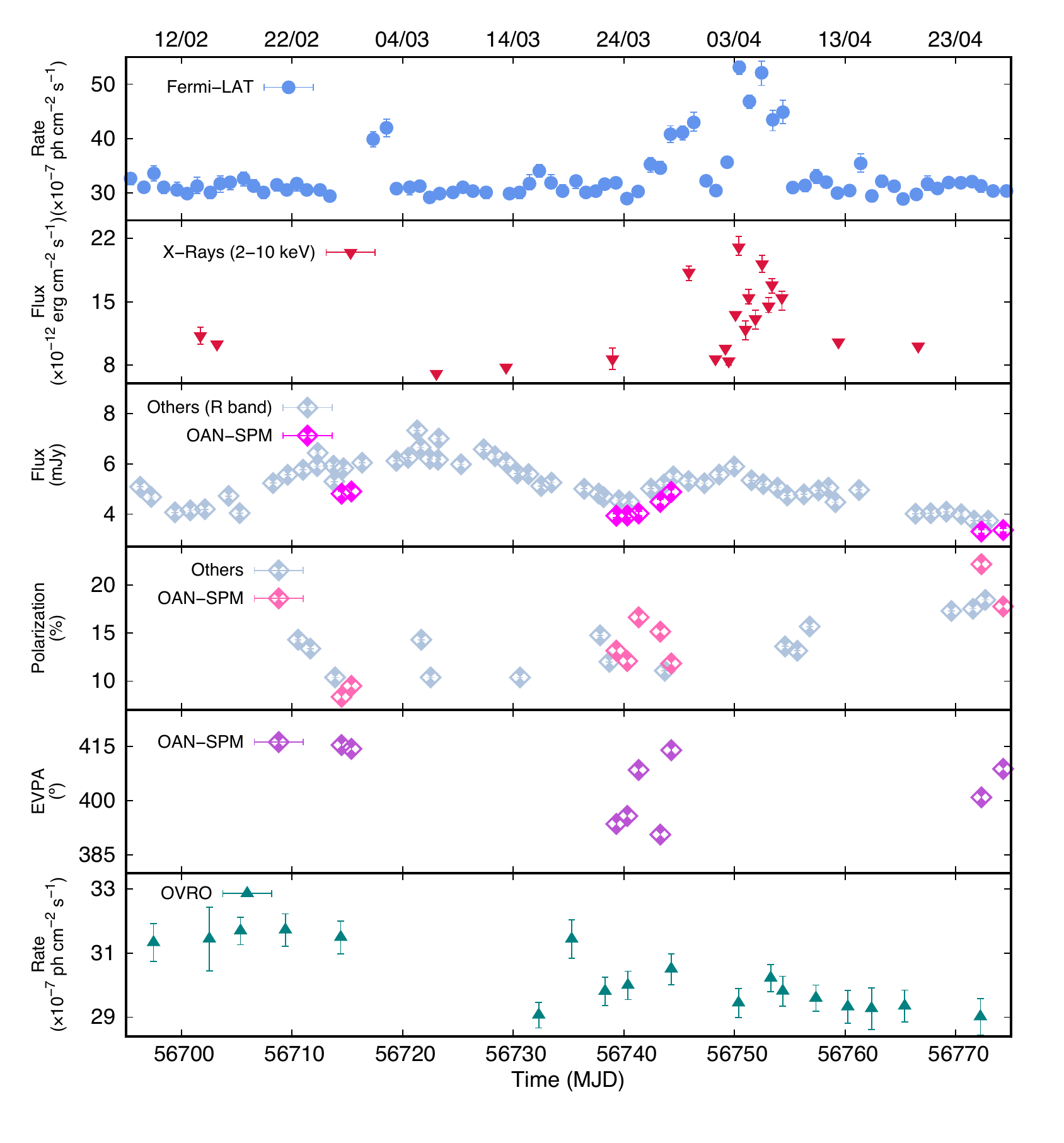}
\caption{Multiwavelength lightcurves of 3C\,279 obtained from  2014 February 07 (MJD 56695) and Abril 28 (MJD 56775) are shown. They include (from top to bottom): $\gamma$-ray flux data above 100 MeV from Fermi-LAT, (0.5 - 5) keV X-ray flux data from Swift-XRT, R-band flux data from OAN-SPM and SMARTS,  P(\%) data from  OAN-SPM and Kanata, EVPA  data from OAN-SPM and radio flux data from OVRO database.}
\label{flare:2014}
\end{figure*}

The multiwavelength lightcurves of 3C\,279 between  2014 February 07 (MJD 56695) and April 28 (MJD 56775) are presented in Figure  \ref{flare:2014}. They include (from top to bottom): $\gamma$-ray flux above 100 MeV (Fermi-LAT), (0.5 - 5) keV X-ray flux collected by Swift-XRT, optical flux in R band measured by OAN-SPM and SMARTS,  polarization degree collected by  OAN-SPM and Kanata, EVPA  measured by OAN-SPM and radio flux detected by OVRO.
During this period, a flaring activity consisting of two multiple sub-structures was present. During the first flare, 3C\,279 showed a moderate activity  in $\gamma$-ray,  X-ray and optical band with random variations in the polarization degree and EVPA. During the second sub-structure, the strongest flare around 2014 Abril 03 (MJD 56750) was detected. It shows  a flaring activity in $\gamma$-ray and X-ray band accompanied with a small increase in optical flux   After the flare (post-flare), the polarization degree increases from 10\% to 20\% in $\Delta t\simeq$1 month and the polarization angle did not show any significant rotation. The previous result indicates that the flaring activity exhibited in 2014 March/April is in agreement with the one-zone SSC model.

The DCF was calculated again in order to estimate  lags among $\gamma$-ray, X-ray and radio bands. During the period between  2014 February  07 (MJD 56695) and April 28 (MJD 56775),  the resulting DCF between  gamma-rays with X-rays  (left) and optical with radio (right) are shown in Figure \ref{dcf}  (lower panels). The DCF between $\gamma$-ray and X-ray bands (left) shows a peak close to zero lag, which points out that both bands are correlated. Nevertheless, the DCF between optical R-band and radio fluxes (right) display a 9.6 days lag, which indicates no correlation between them. Radio and optical-R band fluxes could be produced in different zones.
\subsection{Flaring activity in 2015 June}
\begin{figure*}
\centering
\includegraphics[width=0.95\textwidth]{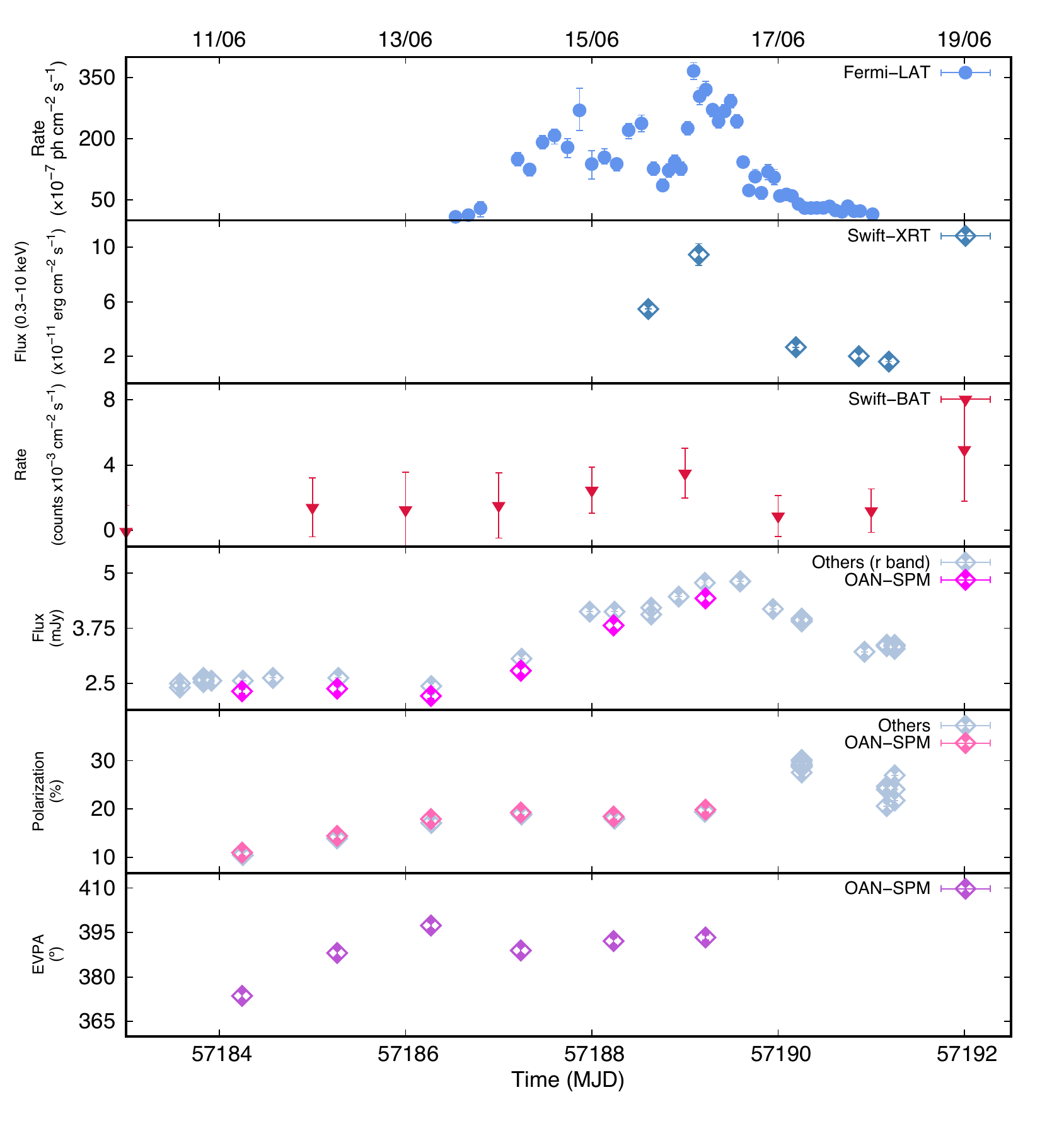}
\caption{Multiwavelength lightcurves of 3C\,279 from 2015 June 13 (MJD 57186) and 19 (MJD 57192) are shown. They include (from top to bottom): $\gamma$-ray flux data above 100 MeV from Fermi-LAT,  X-ray flux data from Swift-BAT and Swift-XRT, R-band and P(\%) data from OAN-SPM and GASP-WEBT and EVPA data from OAN-SPM.}
\label{flare:2015}
\end{figure*}

The multiwavelength lightcurves of 3C\,279 between 2015 June 11 (MJD 57184) and 19 (MJD 57192) are presented in Figure \ref{flare:2015}. They include (from top to bottom): $\gamma$-ray flux above 100 MeV (Fermi-LAT),  X-ray flux collected by Swift-BAT and Swift-XRT, optical flux and polarization degree in R band measured by OAN-SPM and GASP-WEBT and EVPA measured by OAN-SPM.
During the period 2015 June 11 to 17 (MJD 57187 to 57190),  this object exhibited a strong flare in $\gamma$-ray, hard/soft X-ray and optical bands  with a high activity in the polarization degree and random variations of the EVPA. It suggests that the flaring activity was produced by  perpendicular shocks to the directions of axial-symmetric jet which increases  the value of polarization degree without modifying the EVPA \citep{2010Natur.463..919A}.    
\subsection{Flaring activity in 2017 February}
\begin{figure*}
\centering
\includegraphics[width=0.95\textwidth]{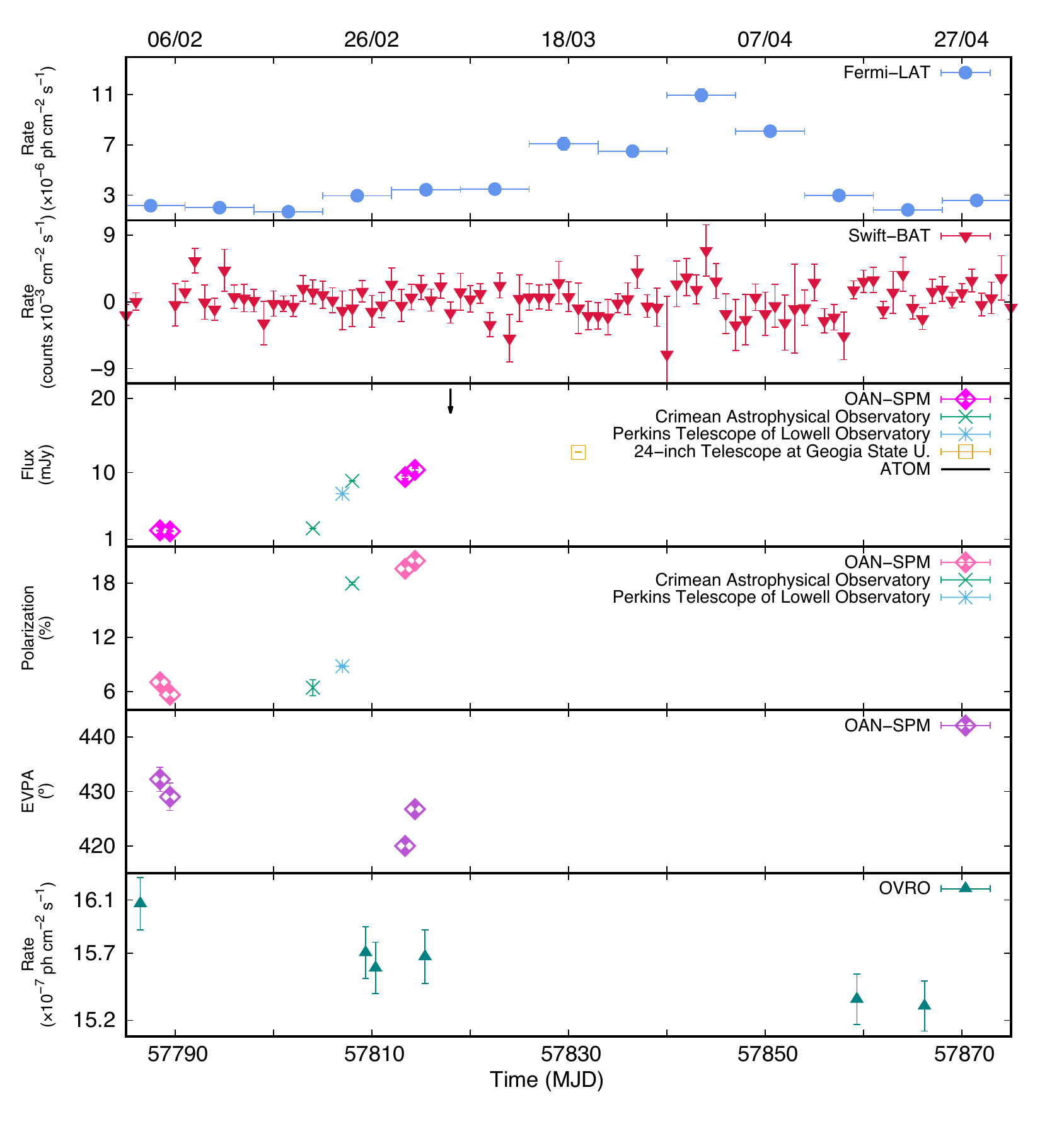}
\caption{3C\,279 lightcurves from 2017 February 01 (MJD 57785) and May 02 (MJD 57875). R- band photopolarimetric data obtained at OAN-SPM are shown. The optical photometric data were complemented with data from the observatories listed in the third panel (from top to bottom). GeV $\gamma$-rays data from Fermi-LAT and Swift-BAT hard X-rays are shown.}
\label{flare:2017}
\end{figure*}

Figure  \ref{flare:2017} shows the R-band photopolarimetric observations along with $\gamma$-ray, hard/soft X-ray, optical and radio data.  The optical flaring activity was monitored in the R-band flux and the polarization degree by AZT-8 telescope of Crimean Astrophysical observatory,  the Perkins telescope of Lowell observatory  \citep{2017ATel10121....1J}, the Automatic telescope for Optical Monitoring  \citep[ATOM;][]{2017ATel10161....1J} and 24-inch telescope at Georgia State University's Hard Labor Creek Observatory \citep{2017ATel10188....1T}.  The flare lasted 90 days and was observed in the optical bands showing a dramatic variation in the  polarization degree while the EVPA remains $\sim$ stable.  A carefully done revision in the literature yields no reports on quasi-simultaneous variability in other wavelengths (in particular, in $\gamma$-ray, X-ray and radio bands). Nevertheless, a change in the optical polarization degree from 6\% to 20\% was observed in the optical band with various telescopes. This atypical optical flare could be explained as the emission of synchrotron radiation produced by a low energy electron population. 
\section{Modeling the broadband emission}
The  adopted scenario to describe the broadband emission in 3C\,279 is the one-zone SSC model with an external radiation component. This external radiation considers as seed photons those coming from the broad line region (BLR) and those produced by the infrared dust (IR).  In the SSC model, electrons are injected  with a double-break power law distribution  $\propto \gamma_e^{-\alpha_1}$ for  $\gamma_{\rm min}<\gamma_e\leq \gamma_{\rm c1}$,  $\gamma_e^{-\alpha_2}$ for $\gamma_{\rm c1}<\gamma_e\leq \gamma_{\rm c2}$ and $\gamma_e^{-\alpha_3}$ for  $\gamma_{\rm c2}<\gamma_e\leq \gamma_{\rm max}$ and  are confined  inside  the emitting zone of radius  $r_{d} \leq \,\frac{\delta_D}{(1+z)}\tau_{v},$  by a magnetic field, that is estimated by equating the synchrotron cooling and the variability timescales \citep[e.g., see][]{2015APh....71....1F, 2016ApJ...830...81F}.    The synchrotron photons  radiated in this zone are  up-scattered to higher energies by the the same electrons via inverse Compton scattering. Considering the number  densities of electrons and protons,  the energy densities carried by electrons $U_e$, protons $U_p$ and the magnetic field $U_B$ can be estimated by {\small $U_e=m_e N_e\langle \gamma_e\rangle$}, {\small $U_p=N_p m_p$} \citep{2015APh....70...54F, 2014ApJ...783...44F} and {\small $U_B=\frac{B^2}{8\pi}$} with $m_p$ the proton mass, $\langle \gamma_e\rangle$ the average electron Lorentz factor and $\Gamma\approx\delta_D$, respectively.  The total jet power  can be calculated through the contributions of electrons, protons and the magnetic field given by $L_{\rm jet}\simeq\pi r_d^2 \Gamma^2 (U_e+U_B+U_p)$.  In order to describe the SEDs in 3C\,279 and obtain the best-fit parameters such as the magnetic fields, number  densities of electrons, Doppler factors and emitting radii, the one-zone SSC model presented in \cite{2017ApJS..232....7F} will be used.  
Given the values found of the energy densities carried by the electrons, protons and  magnetic fields,  the ratios $\lambda_{ij}=\frac{U_i}{U_j}$ indicate that a principle of energy equipartition is given in the jet of 3C\,279.  The gravitational radius of 3C\,279 is $r_g\simeq2.9\times 10^{14}$ cm, and the emitting radius obtained is of the order of $\sim 10^{16}$ cm which is much larger than the gravitational radius\footnote{The supermassive BH mass estimate for 3C\,279 is of $\sim$\,10$^9$M$\odot$, see http://quasar.square7.ch/fqm/1253-055.html}.  
%
The physical properties of the BLR and the IR dust emission have been widely studied. The energy density  of the BLR and the IR dust emission in the coomoving frame is \citep{2012ApJ...754..114H}
\be
u'_{\rm BLR}(r)=\frac{\xi_{\rm BLR} \Gamma^2 L_D}{3\pi r^2_{\rm BLR}(1+\frac{r}{r_{\rm BLR}})^{\beta_{\rm BLR}}}\,,
\ee
and
\be
u'_{\rm IR}(r)=\frac{\xi_{\rm IR} \Gamma^2 L_D }{3\pi r^2_{\rm IR}(1+\frac{r}{r_{\rm IR}})^{\beta_{\rm IR}}}\,,
\ee
respectively. Here, the parameters $\xi_{\rm BLR}$=0.2 and  $\xi_{\rm IR}$=0.4 \citep{2016MNRAS.457.3535Z}  correspond to the fractions of the disk luminosity $L_D$, $r_{\rm BLR}=0.1\left(\frac{L_D}{10^{46}\,{\rm erg\,s^{-1}}}\right)^\frac12\,{\rm pc}$ and $r_{\rm IR}=2.5\left(\frac{L_D}{10^{46}\,{\rm erg\,s^{-1}}}\right)^\frac12\,{\rm pc}$ are the distances where the reprocessing occurs. We use the values of $\beta_{\rm BLR}=3$ \citep{2009ApJ...704...38S} and $\beta_{\rm IR}=4$ \citep{2012ApJ...754..114H}.
\vspace{0.5cm}
\section{Conclusions}
An exhaustive analysis of 9-year optical R-band photopolarimetric data of the FSRQ 3C\,279 from 2008 February 27 to 2017 May 25  was presented.   This source showed in the R-band the  maximum brightness state of $13.68\pm 0.11$ mag ($1.36\pm0.20$ mJy) on 2017 March 02, and  the lowest  brightness state ever recorded of  $18.20\pm 0.87$ mag ($0.16\pm0.03$ mJy) on 2010 June 17, respectively.   The polarization degree was varying between $0.48\pm0.17$\% and $31.65\pm0.77$\% and the EVPA  presented large rotations between $82.98^\circ \pm0.92$ and $446.32^\circ \pm1.95$.   Multi-wavelength analysis covering radio wavelengths to TeV $\gamma$-rays around the flares reported in 2009 February/March, 2011 June, 2014 March/April, 2015 June and 2017 February were analyzed.\\

From the flaring activities the following results are found:

\begin{itemize}
\item In February/March 2009, this source exhibited several $\gamma$-ray, X-ray, optical and radio flaring  events which were accompanied by large variations of EVPA (from  50$^\circ$ to 600$^\circ$) with random variations of the polarization degree.  For instance,  the X-ray light curve exhibited one flare  accompanied with activity in $\gamma$-ray and optical bands and the other one without activity in other wavelengths, i.e. an orphan flare. The minimum variability timescale is $1.49\pm0.09$ days, which corresponds to a size of emitting radius $r_d=(5.03\pm0.03)\times 10^{16}\,{\rm cm}$. The  analysis shows that only when  $\gamma$-ray, X-ray, optical and radio  fluxes decrease   until it reached the quiescent level,  the optical flux and Stokes parameter ${\rm u}$ is strongly correlated. Analysis using DCF suggests that 3C\,279 has multiple emitting regions and that a contribution of secondary pairs $e^\pm$ radiating synchrotron photons from radio wavelengths to X-rays could be present. These pairs are usually generated by the interactions between Fermi-accelerated protons with photons from synchrotron radiation, and by photons from the BLR and/or IR dust emission.

\item In June 2011, 3C\,279 showed two sub-structures with large variations of the optical flux,  the polarization degree and the EVPA. The optical flux is correlated with a high activity in $\gamma$-rays and low activity in X-rays. The DCF suggests contributions from two emitting regions each one with different electron populations. One population with  low electron energies producing  radio wavelengths and the other one with higher energies producing the X-ray to gamma-ray emission.

\item During March/April 2014, a flaring activity consisting of two or multiple different activity states. During the first flare, 3C\,279 showed a moderate activity  in $\gamma$-rays,  X-rays and optical bands with a decay in the polarization degree. During the second flare, the maximum state appeared around 2014 Abril 03 (MJD 56750). During this state, the source showed activity in $\gamma$-rays and X-rays accompanied with a small increase in optical flux  after the flare (post-flare), while the level of $\gamma$-rays, X-rays and optical bands started to decay until it reached the quiescent state level, the polarization degree increased from 10\% to 20\% in $\Delta t\simeq$1 month and the EVPA did not show any significant rotation.   The DCF between optical R-band and radio fluxes (right) show a lag of $\sim$ 7 light-days, which might indicate that both bands were emitted in different zones.  The correlation analysis shows that the flaring activity in March/April 2014 is in agreement with the one-zone SSC model and also with the no lag found between gamma-rays and X-rays (left) during this period.

\item In June 2015,  this object exhibited a strong flare correlated with $\gamma$-rays, hard/soft X-rays and optical bands. These results mean that the observed multi-wavelength variations could be produced in the same emitting region.

\item In February 2017, an intense flare that lasted 90 days was observed in the optical bands with a dramatic variation in  polarization degree, but without reports of activity in the $\gamma$-ray, X-ray and radio bands.  During the flare, the polarization degree varied from 6\% to 20\%, and the polarization angle remained stable. We propose that this flare can be interpreted via synchrotron emission produced by a low energy electron population.

\end{itemize}

The general results of the analysis of the different DCFs obtained with our entirely data-set show that during flaring states there are lags found between different wave-bands. These results suggests that 3C\, 279 has multiple emission regions. It is worth noting that a zero lag was only found between gamma-rays and X-rays around the flaring activity in March/April 2014.
 
The optical variability results obtained with our OAN-SPM long term data-set suggest that in cycle I the source shows a continuum decrease of the optical flux level with a positive correlation with the polarized percentage. The EVPA shows variations that are superimposed to the stable polarized component found to be present in this cycle. 

In cycle II, an outburst induced  the appearance of large rotations of the EVPA.  One scenario is that these rotations could be the result of jet precessions \citep{2019A&A...626A..78B} originated by a  non-axisymmetric changes of the accretion rate.  A second scenario could be that this rotations are produced by an asymmetric toroidal component of the helicoidal magnetic field \citep{2008Natur.452..966M, 2016MNRAS.462.4267J, 2010ApJ...710L.126M}.  Finally, a third possibility is that they are produced by a turbulent magnetic field that randomizes the direction of the polarization vector \citep{1985ApJ...290..627J, 2014ApJ...780...87M}. Based on the long-term behavior observed in the EVPA, the first scenario is preferred, since it explains the smaller value of the viewing angle and the increased flux observed during this cycle.  The presence of the stable component is not detected in this cycle due to the large precession of the jet. In this case, the variable polarized component dominates.  

In cycle III,  the stable component is again detected, but showing a larger value polarization percentage value of $\sim$13\%, i.e. it increased by a factor of $\sim$2. The variable polarization component is less evident in this cycle maybe due to a decrease of the intensity of the precession of the helicoidal jet.   Our analysis indicates that the EVPA has a preferred direction of $\sim$52 deg that is in agreement with the value previously reported by \cite{2018ApJ...858...80R}.  The average continuum flux is found to be higher during this cycle compared with the flux displayed in cycle I.  Finally, it is worth mentioning that a strong optical flare $\sim 10\,{\rm mJy}$ was detected in this cycle without evident counterparts in other bands. During this flare the maximum R-band brightness of 3C\,279 was observed.

The broadband SEDs were fitted within the framework of the one-zone SSC models, including an external inverse Compton scattering (EC) model \citep{2018MNRAS.481.4461F, 2017APh....89...14F}.  
In the one-zone SSC model, an electron population with a simple and also with a double power-law function were used. In addition, an EC model that includes photons from the BLR and from the IR dust were required.  The full set of parameters derived by using our model are: Doppler factor $\delta_D=$ 14 - 18, magnetic field $B=  (0.14 -  2.5)$ G, emitting radius $r_d=$ (1.1 - 2.1) $\times 10^{17}$ cm and electron density $N_e=$(0.1 -  0.26) $\times\, 10^3\, {\rm cm^{-3}}$ with different values of power indexes between 1 and 6.1.  The values reported in our model agree with those found by \cite{2015ApJ...803...15P, 2015ApJ...807...79H,2012ApJ...754..114H, 2014A&A...567A..41A}.  Our best models allowed to estimate that variability timescales are in the range of 3.3 to 4.0 days. 

Current and futures gamma-ray and X-ray telescopes will probe the radiation mechanisms and magnetic field intensity, and morphology of the relativistic jet in blazar 3C279 \citep{2019arXiv190304607R}.
\section*{Acknowledgements}
We thank the anonymous referee for a critical reading of the manuscript and valuable suggestions that helped improve the quality  of  this  paper. N. F.,  E. B.  and M. S.  acknowledge  financial  support  from UNAM-DGAPA-PAPIIT  through  grant IA 102019. E. B. also aknowledge suport from UNAM-DGAPA-PAPIIT grant IN113417. This work is based upon observations carried out at the Observatorio Astron\'omico Nacional on the Sierra San Pedro M\'artir (OAN-SPM), Baja California, Mexico. We gratefully acknowledge Masaaki Hayashida, Vaidehi S. Paliya and Talvikki Hovatta for providing us with part of the data used in this manuscript.   We also  acknowledge  Fermi and Swift public data archives. This research has made use of XRT and LAT data reduction softwares,  the SAO/NASA Astrophysics Data System (ADS)  operated by the Smithsonian Astrophysical Observatory (SAO) under a National Aeronautics and Space Administration (NASA) grant and  the NASA/IPAC Extragalactic Database (NED) operated by the Jet Propulsion Laboratory, California Institute of Technology, under contract with NASA.   This paper has made use of up-to-date SMARTS optical/near-infrared light curves that are available at www.astro.yale.edu/smarts/glast/home.php, and also data collected by the WEBT collaboration and stored in the WEBT archive at the Osservatorio Astrofisico di Torino - INAF (http://www.oato.inaf.it/blazars/webt/); for questions regarding their availability, please contact the WEBT President Massimo Villata ({\tt massimo.villata@inaf.it})
%
%
%
%
\clearpage

\clearpage

\begin{table*}
\begin{center}\renewcommand{\arraystretch}{1.5}\addtolength{\tabcolsep}{4pt}
\caption{R-band photopolarimetric data of  Quasar 3C\,279 from 2008 February 27 to 2017 May 25}\label{all_data}
\begin{tabular}{ c c c c c c c c c}
\hline \hline
 MJD           &     P   &   $\Delta$P  &     PA          & $\Delta$PA  & R$_{\rm mag}$         &  $\Delta$R$_{\rm mag}$ &  $F_{\rm R}$& $\Delta F_{\rm R}$\\
   &  (\%) &   (\%)           &  ($^\circ$)   &   ($^\circ$)   &   &         & (mJy)& (mJy)\\
  (1)            &   (2)    &      (3)          &      (4)         &      (5)           &   (6)    &    (7)            &  (8)   & (9)\\          
 \hline\hline

54523.5190&	16.97&	0.81&	234.12&	1.46&	16.60&	0.21&	0.708&	0.026\\
54524.4873&	15.28&	0.57&	239.88&	1.16&	16.64&	0.20&	0.679&	0.024\\
54525.4487&	20.34&	0.66&	241.69&	1.07&	16.54&	0.19&	0.746&	0.025\\
54526.4790&	17.62&	0.57&	252.32&	1.03&	16.63&	0.20&	0.686&	0.024\\
54534.4600&	10.63&	1.22&	241.05&	3.28&	16.67&	0.24&	0.665&	0.028\\
54535.4390&	13.15&	0.97&	216.91&	2.00&	16.57&	0.22&	0.723&	0.027\\
54536.4204&	16.17&	0.96&	244.05&	1.72&	16.61&	0.22&	0.701&	0.027\\
54540.4072&	13.33&	0.79&	229.31&	1.77&	16.60&	0.21&	0.708&	0.026\\
....& ....& ....& ....& ....& ....& ....& ....& ....\\

\hline
 \end{tabular}
\end{center}
\begin{center}
Data covering the nine years of observations are available in a machine-readable form in the online journal.  Only a small portion of the entire data is shown for guidance. Cols. (2) and (3) show the polarization degree values and their uncertainties. Cols. (4) and (5) show the EVPA values and their uncertainties. Cols. (6) and (7) show the R-band magnitudes and their uncertainties. Cols. (8) and (9) show the optical R-fluxes and their uncertainties.\\
\end{center}
\end{table*}
%
\begin{table*}
\begin{center}\renewcommand{\arraystretch}{2}\addtolength{\tabcolsep}{6pt}
\caption{The maximum, average and minimum values of the R-band photopolarimetric observations}\label{gen_data}
\vspace{0.1cm}
\begin{tabular}{ c c c c }
\hline
\hline
 \normalsize{Parameter}& \normalsize{Max/Date}&\normalsize{Min/Date} & \normalsize{Average} \\
 (1)            &   (2)    &      (3)          &      (4)          \\ 
\hline
\hline
 \normalsize{$F_{\rm R}$(mJy)}                          &  \scriptsize{($10.36\pm0.20$) / (02-03-2017) }        &     \scriptsize{($0.16\pm0.03$) / (17-06-2010)}         &       \scriptsize{$2.09\pm0.01$}  \\
\normalsize{$R_{mag}$}                          &  \scriptsize{($13.68\pm0.11$) / (02-03-2017)}        &     \scriptsize{($18.20\pm0.87$) / (17-06-2010)}         &       \scriptsize{$15.69\pm0.02$}  \\
\normalsize{P(\%)}                         &  \scriptsize{($31.65\pm0.77$) / (17-05-2013)}        &     \scriptsize{($0.48\pm0.17$) / (18-05-2012)}         &       \scriptsize{$14.04\pm0.11$}     \\
\normalsize{$\theta(\degr)$}              &  \scriptsize{($446.32\pm1.95$) / (25-05-2017)}        &     \scriptsize{($82.98\pm0.92$) / (02-06-2011)}         &       \scriptsize{$311.09\pm0.33$}  \\
\hline
\hline
\end{tabular}
\end{center}
\begin{center}
Cols (2),  (3) and (4) show the maximum, minimum and average values of the optical  R-band  flux, magnitude,  polarization degree P(\%) and EVPA ($\theta^{\circ}$). \\
\end{center}
\end{table*}
%
%
\begin{table*}
\begin{center}\renewcommand{\arraystretch}{1.6}\addtolength{\tabcolsep}{6pt}
\caption{Variability of the statistical parameters}\label{stat_cicle}
\begin{tabular}{ c c c c c c c c}
\hline
\hline
\normalsize{Cycle} & \normalsize{Parameter}& \normalsize{Max}&\normalsize{Min} & \normalsize{Average}& \normalsize{Y(\%)}&\normalsize{$\mu(\%)$}& \normalsize{$\cal F $} \\
(1)            &   (2)    &      (3)          &      (4)         &      (5)           &   (6)    &    (7)            &  (8) \\
\hline
\hline
			    & \normalsize{$F_{\rm R}$(mJy)}                          &  \scriptsize{2.13$\pm$0.58}        &     \scriptsize{0.16$\pm$0.03}         &       \scriptsize{$0.71\pm0.01$}      &    \scriptsize{278.92}          &  \scriptsize{0.71}  &  \scriptsize{0.86}                \\
\normalsize{I} & \normalsize{P(\%)}                         &  \scriptsize{29.94$\pm$1.58}        &     \scriptsize{2.71$\pm$0.25}         &       \scriptsize{$14.79\pm0.36$}      &    \scriptsize{182.16}          &  \scriptsize{2.45}  &  \scriptsize{0.83}              \\
			    &\normalsize{$\theta(\degr)$}              &  \scriptsize{336.76$\pm$12.00}        &     \scriptsize{183.88$\pm$4.05}         &       \scriptsize{$253.48\pm0.95$}      &    \scriptsize{59.88}          &  \scriptsize{0.37}    &  \scriptsize{0.29}               \\
\hline
			    & \normalsize{$F_{\rm R}$(mJy)}                          &  \scriptsize{2.13$\pm$0.06}        &     \scriptsize{0.49$\pm$0.05}         &       \scriptsize{$0.98\pm0.01$}      &    \scriptsize{166.11}          &  \scriptsize{0.96}  &  \scriptsize{0.62}                \\
\normalsize{IA} & \normalsize{P(\%)}                         &  \scriptsize{29.94$\pm$1.58}        &     \scriptsize{10.63$\pm$1.22}         &       \scriptsize{$19.24\pm0.54$}      &    \scriptsize{99.28}          &  \scriptsize{2.83}  &  \scriptsize{0.48}               \\
			    &\normalsize{$\theta(\degr)$}              &  \scriptsize{293.57$\pm$2.40}        &     \scriptsize{211.92$\pm$6.14}         &       \scriptsize{$249.52\pm0.78$}      &    \scriptsize{32.51}          &  \scriptsize{0.31}  &  \scriptsize{0.16}                 \\
\hline
			    & \normalsize{$F_{\rm R}$(mJy)}                          &  \scriptsize{1.65$\pm$0.05}        &     \scriptsize{0.41$\pm$0.02}         &       \scriptsize{$0.83\pm0.01$}      &    \scriptsize{149.76}          &  \scriptsize{1.23}   &  \scriptsize{0.60}               \\
\normalsize{IB} & \normalsize{P(\%)}                         &  \scriptsize{17.79$\pm$2.09}        &     \scriptsize{4.27$\pm$1.15}         &       \scriptsize{$10.48\pm0.61$}      &    \scriptsize{124.91}          &  \scriptsize{5.83}     &  \scriptsize{0.61}           \\
			    &\normalsize{$\theta(\degr)$}              &  \scriptsize{322.76$\pm$6.61}        &     \scriptsize{188.65$\pm$2.86}         &       \scriptsize{$269.28\pm1.31$}      &    \scriptsize{49.66}          &  \scriptsize{0.49}  &  \scriptsize{0.26}                \\
\hline
			    & \normalsize{$F_{\rm R}$(mJy)}                          &  \scriptsize{0.60$\pm$0.03}        &     \scriptsize{0.16$\pm$0.03}         &       \scriptsize{$0.30\pm0.01$}      &    \scriptsize{143.87}          &  \scriptsize{1.99}  &  \scriptsize{0.58}               \\
\normalsize{IC} & \normalsize{P(\%)}                         &  \scriptsize{27.05$\pm$3.36}        &     \scriptsize{2.71$\pm$0.25}         &       \scriptsize{$13.85\pm0.72$}      &    \scriptsize{170.88}          &  \scriptsize{5.16}    &  \scriptsize{0.82}              \\
			    &\normalsize{$\theta(\degr)$}              &  \scriptsize{336.73$\pm$12.00}        &     \scriptsize{183.88$\pm$4.05}         &       \scriptsize{$243.86\pm2.40$}      &    \scriptsize{62.25}          &  \scriptsize{0.99}   &  \scriptsize{0.29}               \\
\hline
\hline
			    & \normalsize{$F_{\rm R}$(mJy)}                          &  \scriptsize{5.71$\pm$0.11}        &     \scriptsize{0.87$\pm$0.03}         &       \scriptsize{$2.19\pm0.01$}      &    \scriptsize{221.23}          &  \scriptsize{0.29}  &  \scriptsize{0.74}                \\ 
\normalsize{II} & \normalsize{P(\%)}                         &  \scriptsize{31.65$\pm$0.77}        &     \scriptsize{0.48$\pm$0.11}         &       \scriptsize{$12.51\pm0.10$}      &    \scriptsize{248.96}          &  \scriptsize{0.84}  &  \scriptsize{0.97}              \\
			    &\normalsize{$\theta(\degr)$}              &  \scriptsize{400.00$\pm$0.86}        &     \scriptsize{82.98$\pm$0.92}         &       \scriptsize{$373.71\pm0.48$}      &    \scriptsize{115.82}          &  \scriptsize{0.18}    &  \scriptsize{0.66}               \\
\hline
			    & \normalsize{$F_{\rm R}$(mJy)}                          &  \scriptsize{5.71$\pm$0.11}        &     \scriptsize{0.87$\pm$0.03}         &       \scriptsize{$2.05\pm0.01$}      &    \scriptsize{235.52}          &  \scriptsize{0.51}  &  \scriptsize{0.74}                \\
\normalsize{IIA} & \normalsize{P(\%)}                         &  \scriptsize{26.23$\pm$1.39}        &     \scriptsize{2.59$\pm$0.96}         &       \scriptsize{$13.29\pm0.19$}      &    \scriptsize{176.92}          &  \scriptsize{1.40}  &  \scriptsize{0.82}               \\
			    &\normalsize{$\theta(\degr)$}              &  \scriptsize{334.89$\pm$1.84}        &     \scriptsize{82.98$\pm$0.92}         &       \scriptsize{$201.49\pm0.60$}      &    \scriptsize{125.01}          &  \scriptsize{0.29}  &  \scriptsize{0.65}                 \\
\hline
			    & \normalsize{$F_{\rm R}$(mJy)}                          &  \scriptsize{3.53$\pm$0.09}        &     \scriptsize{1.75$\pm$0.05}         &       \scriptsize{$2.51\pm0.01$}      &    \scriptsize{70.68}          &  \scriptsize{0.46}  &  \scriptsize{0.34}                \\
\normalsize{IIB} & \normalsize{P(\%)}                         &  \scriptsize{18.28$\pm$0.61}        &     \scriptsize{0.48$\pm$0.05}         &       \scriptsize{$8.68\pm0.18$}      &    \scriptsize{204.66}          &  \scriptsize{2.07}  &  \scriptsize{0.95}               \\
			    &\normalsize{$\theta(\degr)$}              &  \scriptsize{364.84$\pm$3.92}        &     \scriptsize{217.36$\pm$1.76}         &       \scriptsize{$300.19\pm1.02$}      &    \scriptsize{49.09}          &  \scriptsize{0.34}  &  \scriptsize{0.25}                 \\
\hline
			    & \normalsize{$F_{\rm R}$(mJy)}                          &  \scriptsize{2.17$\pm$0.05}        &     \scriptsize{1.55$\pm$0.04}         &       \scriptsize{$1.80\pm0.01$}      &    \scriptsize{33.77}          &  \scriptsize{0.55}  &  \scriptsize{0.17}               \\
\normalsize{IIC} & \normalsize{P(\%)}                         &  \scriptsize{31.65$\pm$0.77}        &     \scriptsize{5.43$\pm$0.65}         &       \scriptsize{$18.32\pm0.15$}      &    \scriptsize{142.88}          &  \scriptsize{0.80}    &  \scriptsize{0.71}              \\
			    &\normalsize{$\theta(\degr)$}              &  \scriptsize{400.00$\pm$0.86}        &     \scriptsize{229.68$\pm$0.87}         &       \scriptsize{$338.95\pm0.35$}      &    \scriptsize{50.25}          &  \scriptsize{0.10}   &  \scriptsize{0.27}               \\
\hline
\hline
			    & \normalsize{$F_{\rm R}$(mJy)}                          &  \scriptsize{10.36$\pm$0.19}        &     \scriptsize{1.33$\pm$0.04}         &       \scriptsize{$3.09\pm0.01$}      &    \scriptsize{292.46}          &  \scriptsize{0.33}  &  \scriptsize{0.77}                \\
\normalsize{III} & \normalsize{P(\%)}                         &  \scriptsize{25.93$\pm$0.71}        &     \scriptsize{5.31$\pm$0.38}         &       \scriptsize{$15.69\pm0.11$}      &    \scriptsize{131.23}          &  \scriptsize{0.70}  &  \scriptsize{0.66}              \\
			    &\normalsize{$\theta(\degr)$}              &  \scriptsize{446.32$\pm$1.95}        &     \scriptsize{373.65$\pm$1.27}         &       \scriptsize{$413.71\pm0.23$}      &    \scriptsize{17.55}          &  \scriptsize{0.06}    &  \scriptsize{0.09}               \\
\hline
			    & \normalsize{$F_{\rm R}$(mJy)}                          &  \scriptsize{4.90$\pm$0.14}        &     \scriptsize{1.69$\pm$0.06}         &       \scriptsize{$3.90\pm0.03$}      &    \scriptsize{82.17}          &  \scriptsize{0.70}  &  \scriptsize{0.49}                \\
\normalsize{IIIA} & \normalsize{P(\%)}                         &  \scriptsize{25.13$\pm$1.31}        &     \scriptsize{8.36$\pm$0.41}         &       \scriptsize{$15.36\pm0.21$}      &    \scriptsize{108.45}          &  \scriptsize{1.35}  &  \scriptsize{0.50}               \\
			    &\normalsize{$\theta(\degr)$}              &  \scriptsize{415.31$\pm$1.22}        &     \scriptsize{390.64$\pm$1.88}         &       \scriptsize{$405.67\pm0.37$}      &    \scriptsize{6.03}          &  \scriptsize{0.09}  &  \scriptsize{0.03}                 \\
\hline
			    & \normalsize{$F_{\rm R}$(mJy)}                          &  \scriptsize{4.43$\pm$0.10}        &     \scriptsize{1.33$\pm$0.04}         &       \scriptsize{$2.40\pm0.02$}      &    \scriptsize{128.88}          &  \scriptsize{0.71}   &  \scriptsize{0.54}               \\
\normalsize{IIIB} & \normalsize{P(\%)}                         &  \scriptsize{25.93$\pm$0.71}        &     \scriptsize{8.43$\pm$1.17}         &       \scriptsize{$17.16\pm0.31$}      &    \scriptsize{101.37}          &  \scriptsize{1.79}     &  \scriptsize{0.51}           \\
			    &\normalsize{$\theta(\degr)$}              &  \scriptsize{430.77$\pm$2.90}        &     \scriptsize{373.65$\pm$1.27}         &       \scriptsize{$403.77\pm0.58$}      &    \scriptsize{14.10}          &  \scriptsize{0.14}  &  \scriptsize{0.07}                \\
\hline
			    & \normalsize{$F_{\rm R}$(mJy)}                          &  \scriptsize{3.75$\pm$0.08}        &     \scriptsize{1.47$\pm$0.04}         &       \scriptsize{$2.70\pm0.02$}      &    \scriptsize{84.51}          &  \scriptsize{0.59}  &  \scriptsize{0.44}               \\
\normalsize{IIIC} & \normalsize{P(\%)}                         &  \scriptsize{21.10$\pm$0.59}        &     \scriptsize{5.31$\pm$0.38}         &       \scriptsize{$13.09\pm0.16$}      &    \scriptsize{120.35}          &  \scriptsize{1.23}    &  \scriptsize{0.60}              \\
			    &\normalsize{$\theta(\degr)$}              &  \scriptsize{430.70$\pm$2.90}        &     \scriptsize{388.38$\pm$1.55}         &       \scriptsize{$412.13\pm0.34$}      &    \scriptsize{10.24}          &  \scriptsize{0.08}   &  \scriptsize{0.05}               \\
\hline
			    & \normalsize{$F_{\rm R}$(mJy)}                          &  \scriptsize{10.36$\pm$0.20}        &     \scriptsize{1.89$\pm$0.04}         &       \scriptsize{$3.68\pm0.02$}      &    \scriptsize{230.28}          &  \scriptsize{0.63}  &  \scriptsize{0.69}               \\
\normalsize{IIID} & \normalsize{P(\%)}                         &  \scriptsize{22.11$\pm$0.41}        &     \scriptsize{5.66$\pm$0.48}         &       \scriptsize{$17.23\pm0.13$}      &    \scriptsize{95.31}          &  \scriptsize{0.80}    &  \scriptsize{0.59}              \\
			    &\normalsize{$\theta(\degr)$}              &  \scriptsize{446.32$\pm$1.95}        &     \scriptsize{420.00$\pm$1.40}         &       \scriptsize{$433.19\pm0.43$}      &    \scriptsize{6.03}          &  \scriptsize{0.10}   &  \scriptsize{0.03}               \\

\hline
\hline
\end{tabular}
\end{center}
\begin{center}
Column (1) shows the cycle, col(2) the observables $F_{R}$ or R-band flux, P(\%) is the polarization percentage and the EVPA ($\theta^{\circ}$). Columns (3), (4) and (5) show the maximum, minimum and average values of these observables. Columns (6), (7) and (8) show the amplitude of the variations, the fluctuation index and the fractional variability index.\\
\end{center}
\end{table*}
%
\begin{table*}
\begin{center}\renewcommand{\arraystretch}{1.6}\addtolength{\tabcolsep}{6pt}
\caption{Pearson's correlation coefficients}\label{pearson_year}
\begin{tabular}{ c c c c c c c c}
\hline
\hline
\normalsize{Cycle} &  \normalsize{$F_{\rm R}$-P}&\normalsize{$F_{\rm R}$-$\theta$} & \normalsize{P-$\theta$}& \normalsize{$F_{\rm R}$-u}&\normalsize{$F_{\rm R}$-q}& \normalsize{u-q} \\
(1)            &   (2)    &      (3)          &      (4)         &      (5)           &   (6)    &    (7)       \\
\hline
\hline
\normalsize{ALL} &\scriptsize{-0.52 (3.25$\times10^{-9}$)}          &     \scriptsize{-0.31 (1.5$\times10^{-5}$)}         &       \scriptsize{0.43 (9.7$\times10^{-9}$)}      &    \scriptsize{0.22 (2.4$\times10^{-3}$)}          &  \scriptsize{-1.9$\times10^{-3}$ (0.98)}  &  \scriptsize{0.03 (0.66)}              \\
\hline
\normalsize{I} &\scriptsize{-0.37 (0.01)}          &     \scriptsize{-0.34 (0.02)}         &       \scriptsize{0.29 (0.05)}      &    \scriptsize{-0.42 (3.3$\times10^{-3}$)}          &  \scriptsize{-0.43 (2.6$\times10^{-3}$)}  &  \scriptsize{0.13 (0.38)}              \\
\normalsize{IA} &  \scriptsize{-0.22 (0.40)}        &     \scriptsize{0.03 (0.91)}         &       \scriptsize{0.80 (1.3$\times10^{-4}$)}      &    \scriptsize{-0.94 (3.1$\times10^{-8}$)}          &  \scriptsize{-0.52 (3.3$\times10^{-2}$)}  &  \scriptsize{0.49 (4.8$\times10^{-3}$)}               \\
\normalsize{IB} &  \scriptsize{-0.14 (0.64)}        &     \scriptsize{-0.24 (0.41)}         &       \scriptsize{0.47 (8.8$\times10^{-2}$)}      &    \scriptsize{-0.32 (0.27)}          &  \scriptsize{0.16 (0.59)}     &  \scriptsize{0.27 (0.36)}           \\
\normalsize{IC}  &  \scriptsize{-0.64 (7.4$\times10^{-3}$)}        &     \scriptsize{-0.19 (0.49)}         &       \scriptsize{-0.09 (0.73)}      &    \scriptsize{-0.70 (2.4$\times10^{-3}$)}          &  \scriptsize{-0.01 (0.98)}    &  \scriptsize{0.03 (0.92)}              \\
\hline
\hline
\normalsize{II}  &  \scriptsize{-0.50 (1.3$\times10^{-6}$)}        &     \scriptsize{-0.08 (0.48)}         &       \scriptsize{0.27 (0.01)}      &    \scriptsize{0.26 (0.02)}          &  \scriptsize{-0.12 (0.32)}       &  \scriptsize{0.15 (0.17)}           \\
\normalsize{IIA}  &  \scriptsize{-0.82 (4.1$\times10^{-8}$)}      &     \scriptsize{0.38 (0.04)}         &       \scriptsize{0.59 (6.3$\times10^{-4}$)}      &    \scriptsize{0.64 (1.4$\times10^{-4}$)}          &  \scriptsize{-0.20 (0.31)}          &  \scriptsize{0.03 (0.87)}       \\
\normalsize{IIB}  &  \scriptsize{-0.69 (5.1$\times10^{-6}$)}        &     \scriptsize{-0.27 (0.12)}         &       \scriptsize{-0.04 (0.83)}      &    \scriptsize{0.65 (2.1$\times10^{-5}$)}          &  \scriptsize{-0.24 (0.16)}            &  \scriptsize{-0.36 (0.03)}     \\
\normalsize{IIC}  &  \scriptsize{-0.47 (0.04)}        &     \scriptsize{0.23 (0.34)}         &       \scriptsize{-0.08 (0.76)}      &    \scriptsize{-0.72 (5.1$\times10^{-4}$)}          &  \scriptsize{-0.11 (0.66)}            &  \scriptsize{0.22 (0.37)}     \\
\hline
\hline
\normalsize{III}  &  \scriptsize{-0.31 (0.02)}        &     \scriptsize{-0.13 (0.38)}         &       \scriptsize{0.83 (3.5$\times10^{-10}$)}      &    \scriptsize{0.02 (0.88)}          &  \scriptsize{-0.10 (0.46)}       &  \scriptsize{0.26 (0.05)}           \\
\normalsize{IIIA}  &  \scriptsize{-0.92 (6.7$\times10^{-5}$)}      &     \scriptsize{0.17 (0.62)}         &       \scriptsize{-0.03 (0.94)}      &    \scriptsize{-0.91 (1.2$\times10^{-4}$)}          &  \scriptsize{0.07 (0.83)}          &  \scriptsize{-0.06 (0.87)}       \\
\normalsize{IIIB}  &  \scriptsize{-0.31 (0.25)}        &     \scriptsize{-0.23 (0.39)}         &       \scriptsize{0.98 (2.2$\times10^{-9}$)}      &    \scriptsize{3.2$\times10^{-3}$ (0.99)}          &  \scriptsize{0.57 (0.02)}            &  \scriptsize{-0.11 (0.68)}     \\
\normalsize{IIIC}  &  \scriptsize{-0.59 (0.02)}        &     \scriptsize{0.19 (0.48)}         &       \scriptsize{-0.59 (0.02)}      &    \scriptsize{-0.45 (0.08)}          &  \scriptsize{-0.21 (0.44)}            &  \scriptsize{0.33 (0.22)}     \\
\normalsize{IIID}  &  \scriptsize{-0.59 (0.02)}        &     \scriptsize{-0.34 (0.19)}         &       \scriptsize{0.36 (0.18)}      &    \scriptsize{0.70 (2.4$\times10^{-3}$)}          &  \scriptsize{0.06 (0.84)}         &  \scriptsize{-0.38 (0.15)}        \\
\hline
\hline
\end{tabular}
\end{center}
\begin{center}
Pearson's correlation coefficients among optical flux,  polarization degree, EVPA and the normalized Stokes parameters.  Col (2) Optical flux and P(\%), col (3) optical flux and EVPA, col (4) P(\%) and EVPA. Cols (5) and (6) optical flux and normalized Stokes parameters $u$ and $q$, respectively and col (7) the normalized Stokes parameters $q$ and $u$.  From col (2) to col (7) numbers shown in parenthesis are the estimated p values.\\
\end{center}
\end{table*}
%

\clearpage
%

\begin{landscape}

\begin{table*}
\begin{center}\renewcommand{\arraystretch}{1.5}\addtolength{\tabcolsep}{1pt}
\caption{Leptonic model parameters}
\label{sed_parameters}
\begin{tabular}{ l c c c c c c c c c c}
\hline
\hline
\normalsize{} & \scriptsize{2009 Apr 29 - May 9}&  \scriptsize{2010 Feb 13 - May 3}&  \scriptsize{2011 Feb 8 - Apr 12} &  \scriptsize{2011 Jun 1 - Jun 8} &  \scriptsize{2014 Mar 25 - Apr 2} &  \scriptsize{2015 Jun 16} &  \scriptsize{2015 Jun 16 } \\
\normalsize{} & \scriptsize{(MJD 54950 - MJD 54960)}&  \scriptsize{(MJD 55240 - MJD 55319)}&  \scriptsize{(MJD 55600 - MJD 55663)} &  \scriptsize{(MJD 55713 - MJD 55720)} &  \scriptsize{(MJD 56741 - MJD 56749)} &  \scriptsize{(MJD 57189 - Orbit C )} &  \scriptsize{(MJD 57189 - Orbit D )} \\

\hline
\hline
\multicolumn{2}{c}{\normalsize Fitted quantities} \\
\cline{1-2}
\scriptsize{$\delta_D$} & \scriptsize{18} & \scriptsize{18}&  \scriptsize{15}&  \scriptsize{14}&  \scriptsize{14}&   \scriptsize{16} &  \scriptsize{16}\\
\scriptsize{$B$ (G)} & \scriptsize{0.14}&  \scriptsize{0.14}&  \scriptsize{0.3}&  \scriptsize{0.8}&  \scriptsize{2.5}&  \scriptsize{0.2}&  \scriptsize{0.2} \\
\scriptsize{$r_d$ ($\times 10^{17}$ cm)} & \scriptsize{2.1} & \scriptsize{1.2}   & \scriptsize{1.0}& \scriptsize{1.5}& \scriptsize{1.1}& \scriptsize{1.1} & \scriptsize{1.1} \\
\scriptsize{$N_e$ ($\times 10^{3}$ cm$^{-3}$)} & \scriptsize{2.6} &   \scriptsize{1.7}&  \scriptsize{3.2} &  \scriptsize{1.1} &  \scriptsize{2.3} &   \scriptsize{2.6} &  \scriptsize{2.5}\\
\scriptsize{$\rm p_1$} & \scriptsize{1} &   \scriptsize{1}&  \scriptsize{2.0} &  \scriptsize{2.0} &  \scriptsize{1.7} &    \scriptsize{1.6} &  \scriptsize{1.6}\\
\scriptsize{$\rm p_2$} & \scriptsize{2.5} &   \scriptsize{2.6}&  \scriptsize{3.6} &  \scriptsize{3.4} &  \scriptsize{4.5} &    \scriptsize{3.8} &  \scriptsize{3.6}\\
\scriptsize{$\rm p_3$} & \scriptsize{6.1} &   \scriptsize{6.1}&  \scriptsize{-} &  \scriptsize{-} &  \scriptsize{-} &  \scriptsize{-} &  \scriptsize{-}\\\hline
\multicolumn{2}{c}{\normalsize Derived quantities} \\
\cline{1-2}
\scriptsize{$\gamma_{c1}$ ($\times 10^2$) } & \scriptsize{8.1}&  \scriptsize{8.2}&  \scriptsize{6.0}&  \scriptsize{6.1}&  \scriptsize{6.9}&  \scriptsize{7.6}&  \scriptsize{7.6} \\
\scriptsize{$\gamma_{c2}$ ($\times 10^4$) } & \scriptsize{6.5}&  \scriptsize{6.6}&  \scriptsize{-}&  \scriptsize{-}&  \scriptsize{-}&    \scriptsize{-}&  \scriptsize{-} \\
\scriptsize{$U_e$\,\, ($\times 10^{-1}$)}   & \scriptsize{0.9}&  \scriptsize{0.4}&  \scriptsize{0.4}&  \scriptsize{0.2}&  \scriptsize{0.3}&    \scriptsize{0.3}&  \scriptsize{0.3} \\
\scriptsize{$U_B$\,\, ($\times 10^{-3}$)}   & \scriptsize{0.8}&  \scriptsize{0.8}&  \scriptsize{3.6}&  \scriptsize{25.4}&  \scriptsize{284.7}&   \scriptsize{1.6}&  \scriptsize{1.6} \\
\scriptsize{$L_e$  ($\times 10^{46}$ erg/s)} & \scriptsize{8.6}&  \scriptsize{1.6}&  \scriptsize{0.9}&  \scriptsize{0.6}&  \scriptsize{0.6}&   \scriptsize{1.0}&  \scriptsize{1.0} \\
\scriptsize{$L_B$  ($\times 10^{45}$ erg/s)} & \scriptsize{1.1}&  \scriptsize{0.3}&  \scriptsize{0.7}&  \scriptsize{10.1}&  \scriptsize{55.5}&  \scriptsize{0.5} &  \scriptsize{0.5} \\
\scriptsize{$L_p$  ($\times 10^{48}$ erg/s)} & \scriptsize{5.2}&  \scriptsize{1.1}&  \scriptsize{1.0}&  \scriptsize{0.6}&  \scriptsize{0.7}&   \scriptsize{1.1}&  \scriptsize{1.1} \\
\scriptsize{$L_{\rm jet}$  ($\times 10^{48}$ erg/s)} & \scriptsize{5.3}&  \scriptsize{1.1}&  \scriptsize{1.0}&  \scriptsize{0.7}&  \scriptsize{0.8}&   \scriptsize{1.1}&  \scriptsize{1.1}  \\
\hline
\hline
\end{tabular}
\end{center}

{\scriptsize 
{\bf Note.}  The viewing angle is assumed as 2$^\circ$, the minimum electron Lorentz factor $\gamma_{m}=10$ and the typical temperature of the torus 850 ${\bf K}$.\\
{\bf The fitted quantities}: $\delta_D$ is the Doppler factor,  $B$ is the magnetic field intensity,  $r_d$ is the radius of the emitting region,  $N_e$ is the electron number density,   ${\rm p_1}$, ${\rm p_2}$ and ${\rm p_3}$ are the lower, medium and higher power indexes of the electron population, respectively,   $\gamma_{c1}$,  $\gamma_{c2}$ and $\gamma_{e,max}$ are the electron Lorentz factors for breaks (1 and 2) and maximum, respectively.\\
{\bf The derived quantities}:  $U_e$ and $U_B$ are the densities carried by electrons and magnetic field, respectively,  $L_e$,  $L_B$ and $L_p$ are the electron, magnetic field and proton luminosities, and $L_{\rm jet}$ is the total jet power, respectively.}

\end{table*}




\clearpage

\end{landscape}

%
%
%


%
%
\begin{figure*}
\centering
\includegraphics[width=19 cm , height=23 cm]{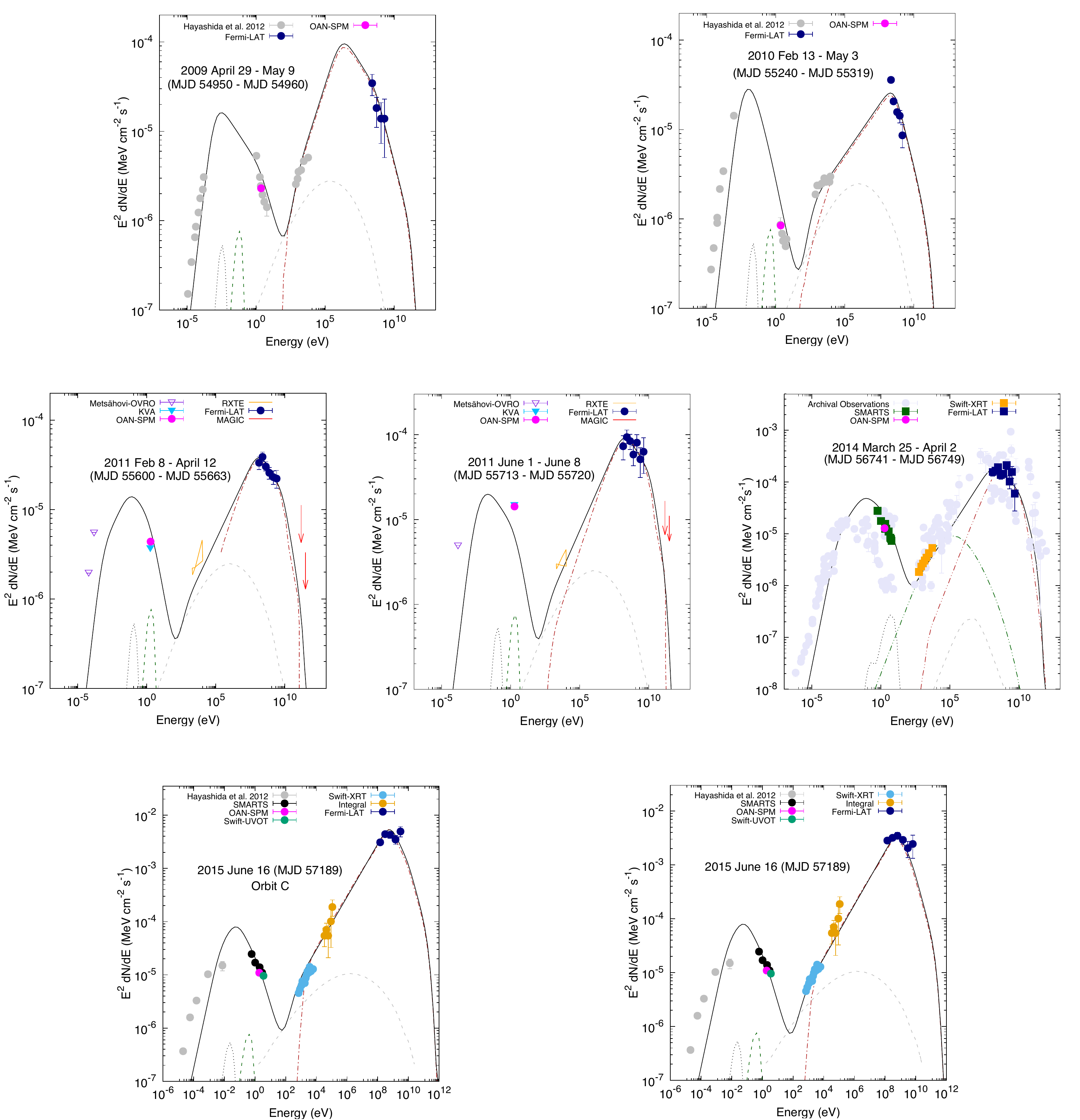}
\caption{The one-zone SSC model with an external radiation component (seed photons from the BLR and IR dust) radiation  was used to fit the SEDs of 3C\,279 during the flares observed in May 2008, March 2010, and the quiescent states from 2008 August 05 (MJD 54683) to  2009 June 18 (MJD 55000) and from 2010 November 16 (MJD 55516) to 2012 June 28 (MJD 56106).  The best-fit parameters are reported in Table 5.}
\label{periods}
\end{figure*}

\end{document}